\newcommand{\gr}{$\gamma$-ray}
\newcommand{\lsim}{{\lower.5ex\hbox{$\; \buildrel < \over \sim \;$}}}
\newcommand{\gsim}{{\lower.5ex\hbox{$\; \buildrel > \over \sim \;$}}}
\newcommand{\nupeak}{$\nu_{\rm peak}$}
\newcommand{\nufnu}{$\nu$f$({\nu})$}

\documentclass[fleqn,usenatbib]{mnras}

\usepackage{newtxtext,newtxmath}

\usepackage{longtable}

\usepackage[T1]{fontenc}

\DeclareRobustCommand{\VAN}[3]{#2}
\let\VANthebibliography\thebibliography
\def\thebibliography{\DeclareRobustCommand{\VAN}[3]{##3}\VANthebibliography}

\usepackage{graphicx}	
\usepackage{amsmath}	
\usepackage{amssymb}	

\title[X-ray data of blazars frequently observed by Swift]{X-ray spectra, light-curves and SEDs of blazars frequently observed by Swift}

\author[Giommi et al.]{
 Paolo Giommi$^{1,2,3,4}$,\thanks{E-mail: giommipaolo@gmail.com}
 M. Perri   $^{5,6}$,
 M. Capalbi $^{7}$,
 V. D'Elia $^{5,8}$, 
 U. Barres de Almeida $^{9}$, 
 \newauthor  
 C.H. Brandt  $^{10,4,9}$,
 A.M.T. Pollock $^{11}$,
F. Arneodo $^{12}$,
 A. Di Giovanni $^{12}$,
 Y. L. Chang  $^{13}$,
 \newauthor 
 O. Civitarese $^{14,15}$,
 M. De Angelis $^{8}$,
 C. Leto  $^{8}$,
 F. Verrecchia  $^{5,6}$,
 N. Ricard $^{16}$,
 S. Di Pippo $^{16}$,
 \newauthor
 R. Middei $^{5}$,
 A. V. Penacchioni $^{14}$,
 R. Ruffini $^{4,17}$,
 N. Sahakyan $^{18,4,17}$,
 D. Israyelyan $^{18}$,
 S. Turriziani $^{19}$
\\
\\
$^{1}$Institute for Advanced Study, Technische Universit{\"a}t M{\"u}nchen, Lichtenbergstrasse 2a, D-85748 Garching bei M\"unchen, Germany\\
$^{2}$ Associated to Italian Space Agency, ASI, via del Politecnico snc, 00133 Roma, Italy\\
$^{3}$ Center for Astro, Particle and Planetary Physics (CAP3), New York University Abu Dhabi, PO Box 129188 Abu Dhabi, United Arab Emirates;\\
$^{4}$ICRANet, P.zza della Repubblica 10, 65122, Pescara, Italy\\
$^{5}$ Space Science Data Center, SSDC, ASI, via del Politecnico snc, 00133 Roma, Italy\\
$^{6}$ INAF - Osservatorio Astronomico di Roma, via di Frascati 33, I-00078 Monteporzio Catone, Italy\\
$^{7}$ INAF - Istituto di Astrofisica Spaziale e Fisica Cosmica di Palermo, via Ugo La Malfa 153, I-90146 Palermo, Italy\\
$^{8}$ Italian Space Agency, ASI, via del Politecnico snc, 00133 Roma, Italy\\
$^{9}$Centro Brasileiro de Pesquisas F\'isicas, Rua Dr. Xavier Sigaud 150, 22290-180, Rio de Janeiro, Brazil\\
$^{10}$ Jacobs University, Physics and Earth Sciences, Campus Ring 1, 28759, Bremen, Germany\\
$^{11}$ Department of Physics and Astronomy, University of Sheffield, Hounsfield Road, Sheffield S3 7RH, England\\
$^{12}$ New York University Abu Dhabi, Abu Dhabi, UAE\\
$^{13}$ Tsung-Dao Lee Institute, Shanghai Jiao Tong University, 800 Dongchuan RD. Minhang District, Shanghai, China\\
$^{14}$ Institute of Physics. IFLP-CONICET. diag 113 e/63-64.(1900) La Plata. Argentina\\
$^{15}$ Department of Physics. University of La Plata. 49 and 115.C.C.67 (1900) La Plata, Argentina\\
$^{16}$ United Nations Office for Outer Space Affairs, UNOOSA, Vienna, Austria\\
$^{17}$ ICRA, Dipartimento di Fisica, Sapienza Universita` di Roma, P.le Aldo Moro 5, 00185 Rome, Italy\\
$^{18}$ ICRANet-Armenia, Marshall Baghramian Avenue 24a, Yerevan 0019, Armenia\\
$^{19}$ Physics Department, Gubkin Russian State University (National Research University), 65 Leninsky Prospekt, Moscow, 119991, Russian Federation
}

\date{Accepted XXX. Received YYY; in original form ZZZ}

\pubyear{2021}

\begin{document}
\label{firstpage}
\pagerange{\pageref{firstpage}--\pageref{lastpage}}
\maketitle

\begin{abstract}
Blazars research is one of the hot topics of contemporary extragalactic astrophysics. That is because these sources are the most abundant type of extragalactic \gr\, sources and are suspected to play a central role in multi-messenger astrophysics.
We have used swift$\_$xrtproc, a tool to carry out an accurate spectral and photometric analysis of the Swift-XRT data of all blazars observed by Swift at least 50 times between December 2004 and the end of 2020.
We present a database of X-ray spectra, best-fit parameter values, count-rates and flux estimations in several energy bands of over 31,000 X-ray observations and single snapshots of 65 blazars.
The results of the X-ray analysis have been combined with other multi-frequency archival data to assemble the broad-band Spectral Energy Distributions (SEDs) and the long-term lightcurves of all sources in the sample.
Our study shows that large X-ray luminosity variability on different timescales is present in all objects. Spectral changes are also frequently observed with a "harder-when-brighter" or "softer-when-brighter" behaviour depending on the SED type of the blazars. The peak energy of the synchrotron component (\nupeak\,) in the SED of HBL blazars, estimated from the log-parabolic shape of their X-ray spectra, also exhibits very large changes in the same source, spanning a range of over two orders of magnitude in Mrk421 and Mrk501, the objects with the best data sets in our sample.
\end{abstract}

\begin{keywords}
galaxies:active – quasars:general – X-rays:galaxies  -- Galaxies: BL Lacertae objects: -- Methods: data analysis -- Astronomical data bases:catalogues
\end{keywords}


\section{Introduction}

Blazars are the most powerful non-explosive sources in the Universe. What makes these sources special compared to other types of Active Galactic Nuclei \citep[AGN,][]{AGNReview} is that their electromagnetic emission is dominated by non-thermal radiation that is generated within a jet that moves away from the central supermassive black hole at relativistic speeds and points in the direction of the Earth \citep[see e.g.][]{Blandford1978,Urry1995,AGNReview}. 

Blazars are sub-classified as Flat-Spectrum Radio Quasars (FSRQs) and BL Lacertae objects (or BL Lacs), based on their optical spectra: FSRQs show broad emission lines like radio-quiet Quasi-Stellar Objects (QSOs), while BL Lacs display at most very weak emission lines, and in several cases are completely featureless \citep[e.g.][]{2014A&ARv..22...73F}.

The radio to \gr\, Spectral Energy Distribution of blazars (SED, a plot of the  \nufnu\  flux as a function of frequency $\nu$) always displays a "double humped" shape \citep[see e.g. ][]{,Abdo2010,GiommiPlanck}. 
The low-energy hump, peaking between the far IR and the X-ray band, is generally attributed to synchrotron radiation produced by relativistic particles moving in a magnetic field inside the jet. The second component, which spans from the X-rays to the \gr\ band, is usually explained as inverse Compton scattering of the electrons against synchrotron generated photons, or other photon field.
Depending on whether the peak of the synchrotron part of the SED (\nupeak\,) is located at low, intermediate or high frequencies (or energy) blazars have been classified as objects of the LBL, IBL or HBL type\footnote{LBL: \nupeak~$<10^{14}$~Hz;  IBL: $10^{14}$~Hz$<$ \nupeak~$<10^{15}$~Hz; HBL: \nupeak~$>10^{15}$~Hz.}, originally for BL Lac objects only \citep{padovani1995}. This classification was subsequently extended to FSRQs by \cite{Abdo_2010} who used the LSP, ISP and HSP coding. In this paper we adopt the original LBL, IBL and HBL codification for all blazars types.

Blazars are central to today's extragalactic high-energy astrophysics as they are by far the most common type of \gr\, sources at high Galactic latitudes \citep{4FGL,4FGLDR2}, and are expected to be abundantly detected in the very high-energy (E$\gsim$50 GeV) \gr\ sky
 that  will  soon  be  surveyed  by  a new  generation  of  VHE  observatories  such  as  CTA \citep{CTAbook} and LHAASO \citep{lamura2020}\footnote{http://english.ihep.cas.cn/lhaaso}.
Blazars have also been proposed to play a key role in the emerging field of multi-messenger astrophysics \citep[e.g.][]{Mannheim1995,Resconi2017}, and their relevance in this area is growing after the association of the object TXS0506+056, and possibly several other blazars, with some IceCube astrophysical high-energy neutrinos \citep[e.g.][]{neutrino,Dissecting,Giommidissecting}.

As part of the Open Universe initiative \citep{GiommiOU}  we have recently started a series of activities aiming at the generation of transparent scientific products from multi-frequency data obtained by the Neil Gehrels Swift observatory \citep[][hereafter Swift]{swift} and other astronomy satellites.
One such program, Open Universe for blazars \footnote{https://sites.google.com/view/ou4blazars}, is dedicated to the class of blazars.

Open Universe is an initiative of the United Nations Office for Outer Space Affairs (UNOOSA) with the objective of making astronomy and space science data more openly available, easily discoverable, free of bureaucratic, administrative or technical barriers, and therefore usable by the widest possible community, from professional researchers to all people interested in space science.
In addition to generating impact on education and capacity building for the XXI Century, one of the main goals of Open Universe is to increase the productivity of space research. By doing so, it aims to contribute to the democratisation of space science and to the achievement of some of the United Nations Sustainable Development Goals (SDGs).
The Initiative, initially proposed by Italy to the Committee On the Peaceful Uses of Outer Space (COPUOS) in 2016, is now under implementation within UNOOSA, with the contribution of a number of scientists and institutions from various countries.

In an earlier paper \citep[][hereafter Paper I]{paper1} we presented a new generation of astronomical products for all the catalogued blazars observed by Swift-XRT during its first 14 years of activity. 
In this paper we present a detailed X-ray spectral, imaging and timing analysis of all the observations and single snapshots\footnote{A Swift observation snapshot is the time interval spent continuously observing a target. A complete observation is composed of one or more snapshots sharing the same observation ID} of a sample of 65 blazars that have been observed by Swift more than 50 times over a period of 16 years, from launch to the end of 2020. The resulting science-ready data products are used to compile the SED and X-ray lightcurve of each source in the sample. 
A preliminary version of this project was presented by \cite{Swift10years} on the occasion of the tenth anniversary of the launch of the Swift mission.
All the results are available through the Open Universe platform, the ASI SSDC, the Virtual Observatory (VO), and through other data release methods. In particular, all the SED data points and lightcurves are accessible through the VOU-blazars tool \citep{VOU-Blazars}.

This work is meant to be a contribution to the goal of creating a high-transparency database, dedicated to blazars, based on the most advanced principles of open data access and behavioural insight approaches\footnote{https://www.oecd.org/gov/regulatory-policy/behavioural-insights.htm}.

The paper is structured as follows. In Section 2 we introduce the sample, in Section 3 we describe in detail the data analysis, in Section 4 we present the results of the  analysis of the XRT data, together with the broad band spectral energy distributions and X-ray lightcurves. In Section 5 we discuss the results.

\section{Blazars frequently observed by Swift-XRT}\label{sample}

\begin{figure}
\centering
\includegraphics[width=8.5cm]{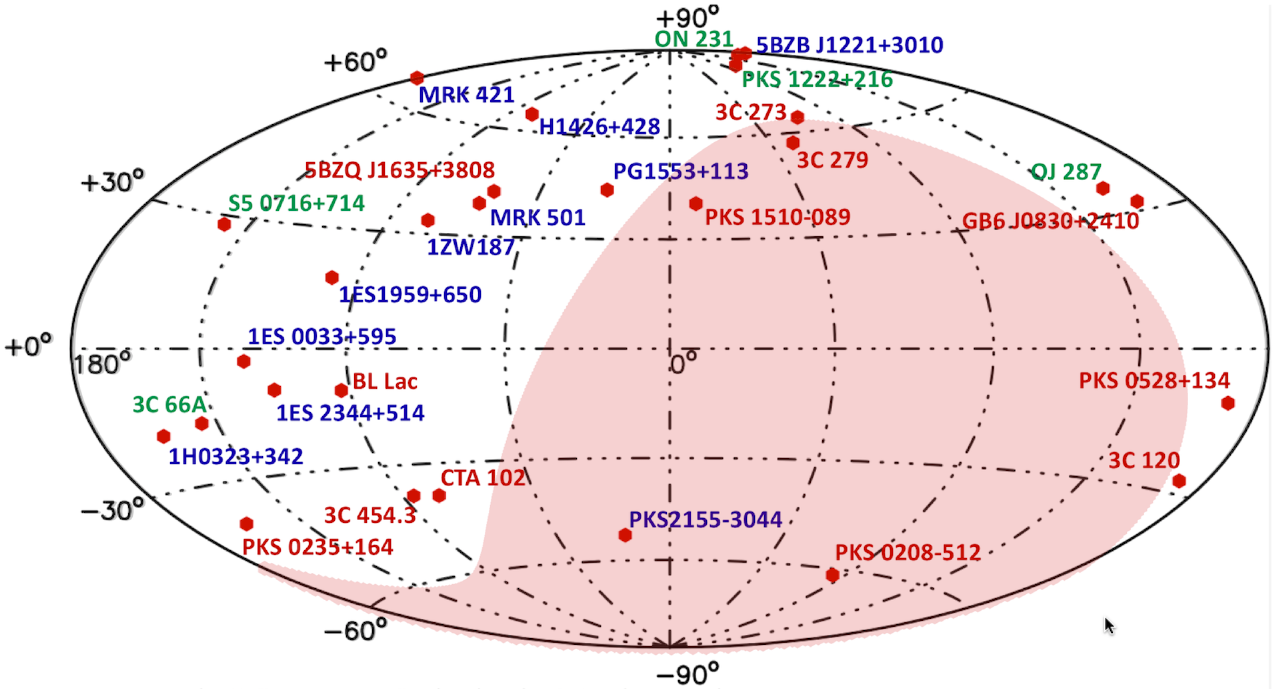}
\caption{Hammer-Aitoff plot in Galactic coordinates showing the position of blazars  that have been observed more than 100 times by Swift. The names of HBL sources appear in blue color, while
those of IBLs and LBLs are green and red respectively. The light red area highlights the part of the sky south of the equator, illustrating how most of the blazars frequently observed by Swift are in the northern hemisphere.}
\label{fig:AitoffDistribution}       
\end{figure}

Although specifically designed for Gamma Ray Burst science, 
Swift has proven to be an extremely effective multi-purpose multi-frequency observatory. A large number of bright and highly variable X-ray sources have been observed many times between launch in November 2004, and the end of 2020. The list of blazars that have been observed at least 50 times in this period with the X-ray telescope \citep[XRT,][]{Burrows2005} is given in Table \ref{TheSample}, where
column 1 is the source common or historical name, column 2 is the name of the object in the BZCAT \citep{Massaro2015} or the 3HSP \citep{3HSP} catalogues, or following the IAU denomination in case the object is not listed in these two catalogues,
column 3 gives the SED classification of the blazar (LBL, IBL or HBL),
column 4 gives the number of XRT observations with exposure larger than 200 seconds in Photon Counting (PC) or Windowed Timing (WT) mode  \citep[see][for details of the readout modes]{Burrows2005}, column 5 gives the ratio between the minimum and maximum flux observed at 1keV. 
The list includes 65 objects, 24 of which 
are HBLs, 12 are IBLs and 29 are LBLs.
The sky distribution of the subsample of sources observed more than 100 times shown in Fig.\ref{fig:AitoffDistribution}
shows that the blazars most pointed by Swift are mostly located in the northern sky. 
 
Although the data presented in this paper is probably the largest available set of homogeneous X-ray measurements of blazars, it is far from being a collection of observations carried out at random times on a sample of randomly selected sources that would be needed for an unbiased view of blazars. Any statistical consideration based on the results presented here should therefore take into account of possible significant selection biases.

\begin{scriptsize}
\begin{table}
\setlength{\tabcolsep}{1.pt}
\begin{center}
\caption{The sample of blazars observed by Swift XRT in PC or WT mode more than 50 times from launch in November 2004 to November 2020. See Sect. \ref{sample} for a description of the columns}
\begin{tabular}{llccc}
\hline\hline
Common or  & 5BZB or 3HSP  & SED & Obs. & f$_{\rm max}$/f$_{\rm min}$ \\
discovery name &IAU denomination &type &  with good & @{\rm 1keV} \\
& & & spectra& \\
~~~~~~~~~~~(1) &  ~~~~~~~~~~~(2)& (3) & (4) & (5) \\
\hline
1ES0033+595     &	   3HSPJ003552.6+595004  	& HBL & 141  & 8.3 \\
PKS0208$-$512   &	   5BZUJ0210-5101        & LBL	& 161  &31.7 \\
3C66A          	&	   5BZBJ0222+4302        & IBL	& 109  &16.2	\\
1ES0229+200    	&	   3HSPJ023248.6+201717  & HBL	&  80  & 3.8 \\
PKS0235+164     &	   5BZBJ0238+1636        &LBL	& 185  & 172.8 \\
1H0323+342     	&	   5BZUJ0324+3410        &IBL	& 147  &6.8	\\
1ES0414+009    	&	   3HSPJ041652.5+010524  & HBL	& 51  &	4.3 \\
3C120          	&	   5BZUJ0433+0521        &LBL	& 176  &	4.3 \\
PKS0506-61     	&	   5BZQJ0506-6109        &LBL	& 59  &	4.0 \\
TXS0506+056  &  5BZBJ0509+0541   &IBL   & 82  & 13.4 \\
GB6J0521+2113   &	   3HSPJ052146.0+211251  & HBL	& 81  &	110.4 \\
5BZQJ0525-4557  &      5BZQJ0525-4557        & LBL  & 51  & 7.2  \\ 
PKS0528+134    	&	   5BZQJ0530+1331        &LBL	& 134  &	25.2  \\
RXSJ05439-5532  &	   3HSPJ054357.2-553208  &HBL	&77  &	7.9 \\
PKS0548-322    	&	   3HSPJ055040.6-321616  &HBL	& 63  &	3.8 \\
PKS0637-752   	&	   5BZQJ0635-7516        &LBL	&67 &	3.5 \\
1ES0647+250    	&	   3HSPJ065046.5+250360  &HBL	& 60  &	8.4 \\
5BZBJ0700-6610  &      5BZBJ0700-6610        & IBL  & 65 & 7.9 \\
EXO0706.1+5913  &	   3HSPJ071030.1+590821  &HBL	&64	&	5.3 \\
S50716+714     	&	   5BZBJ0721+7120        &IBL	& 329  & 69.6\\
3FGLJ0730.5-6606&	   3HSPJ073049.5-660219  &HBL	& 64 &24.6	\\
GB6J0830+2410   &	   5BZQJ0830+2410        &LBL	&104 &48.9 \\
S50836+71   	&	   5BZQJ0841+7053        &LBL	&77  &	5.5 \\
GB6J0849+5108   &	   5BZUJ0849+5108        &LBL	&56  &	20.9 \\
OJ287          	&	   5BZBJ0854+2006        &IBL	&670 & 122.5	\\
PKS0921-213    	&	   5BZUJ0923-2135        &LBL	&95  &	5.0 \\
S40954+658      &	   5BZBJ0958+6533        &LBL	& 58 &57.0	\\
1ES1011+496    	&	   3HSPJ101504.1+492601  &HBL	& 48 &12.5 \\
Mrk421 	        &      3HSPJ110427.3+381232  &HBL	&1163 &	107.4 \\
PKS1130+009     &   5BZQJ1133+0040   & LBL  & 58  &  13.9  \\
GB6J1159+2914   &	   5BZQJ1159+2914        & LBL & 75 &17.3 \\
MS1207.9+3945  	&	   3HSPJ121026.6+392908  &HBL	& 411 & 6.5 \\
1ES1218+304    	&	   3HSPJ122122.0+301037  &HBL	& 113 & 12.5 \\
ON231           &	   5BZBJ1221+2813        &IBL  & 120 & 82.0 \\
PKS1222+216    	&	   5BZQJ1224+2122        &IBL  & 113 & 10.5 \\
3C273          	&	   5BZQJ1229+0203        &LBL  &326  & 10.2 \\
3HSP J123800+263553 &   3HSP J123800+263553  & HBL & 83 & 5.5  \\
3C279          	&	   5BZQJ1256-0547        & LBL	&421  & 11.4 \\
PKS1406-076    	&	   5BZQJ1408-0752        & LBL	&82  & 8.6  \\
PKS1424+240    	&	   5BZBJ1427+2348        & LBL &85 & 16.3 \\
PKS1424-41     	&	   5BZQJ1427-4206        & LBL &63 &	7.4 \\
H1426+428      	&	   3HSPJ142832.6+424021  &HBL & 179 & 3.8 \\
PKS1502+106 	&	   5BZQJ1504+1029  & LBL & 105 &	8.9 \\
PKS1510-08     	&	   5BZQJ1512-0905  & IBL &267  &	6.8 \\
1H1515+660      &      3HSPJ151747.6+652523  & HBL	& 54 &	5.7 \\
4FGLJ1544.3-0649 &      J154419.7-064916  & HBL & 50 & 34.7 \\
PG1553+113     	&	   3HSPJ155543.1+111124  &HBL	& 333 & 21.6 \\
PKS1622-297     	&	   5BZUJ1626-2951  & LBL & 85 &	122.7 \\
B31633+382      &	   5BZQJ1635+3808   & LBL &144 & 27.2 \\
Mrk501         	&	   3HSPJ165352.2+394537  &HBL	&738  &16.6 \\
IZW187      	&	   3HSPJ172818.6+501311  &HBL	&160  & 13.1 \\
PKS1730-130    	&	   5BZQJ1733-1304   & LBL & 93 & 14.0 \\
S41749+701    	&	   5BZBJ1748+7005    &	IBL & 57 &	12.2 \\
S51803+784     	&	   5BZBJ1800+7828    & LBL	&72  &	6.0 \\
3C371          	&	   5BZBJ1806+6949    & IBL  &62  & 10.0 \\
EXO1811.7+3143  &	   3HSPJ181335.2+314418  &HBL & 57 & 172.0 \\
2E1823.3+5649   &	   5BZBJ1824+5651   & LBL & 70  & 7.5 \\
4C+56.27 &   5BZB J1824+5651  & LBL  & 66 & 7.5 \\
PKS1830-211    	&	   5BZQJ1833-2103  & LBL &72 & 3.8 \\
\hline\hline
\label{TheSample}
\end{tabular}
\end{center}
\end{table}
\end{scriptsize}

\begin{scriptsize}
\begin{table}
\setlength{\tabcolsep}{2.0pt}
\begin{center}
\setcounter{table}{0}
\caption{continued}
\begin{tabular}{llccc}
\hline\hline
Common or  & 5BZB or 3HSP  & SED & Obs. & f$_{\rm max}$/f$_{\rm min}$ \\
discovery name &IAU denomination &type &  with good & @{\rm 1keV} \\
& & & spectra& \\
~~~~~~~~~~~(1) &  ~~~~~~~~~~~(2)& (3) & (4) & (5) \\
\hline
1ES1959+650    	& 3HSPJ195959.8+650855  & HBL & 487 & 20.0 \\
PKS2155-304    	&	   3HSPJ215852.1-301332  &HBL	&240 &57.3 \\
BLLac          	&	   5BZBJ2202+4216        &IBL	&545 & 206.6 \\
CTA102         	&	   5BZQJ2232+1143        & LBL & 138 & 20.8 \\
3C454.3        	&	   5BZQJ2253+1608        & LBL	&414 & 24.4 \\
1ES2344+514    	&	   3HSPJ234704.8+514218  &HBL	&184 & 10.3 \\
\hline\hline
\label{TheSampleContinued}
\end{tabular}
\end{center}
\end{table}
\end{scriptsize}

\begin{figure*}
\centering
\includegraphics[width=17.5cm]{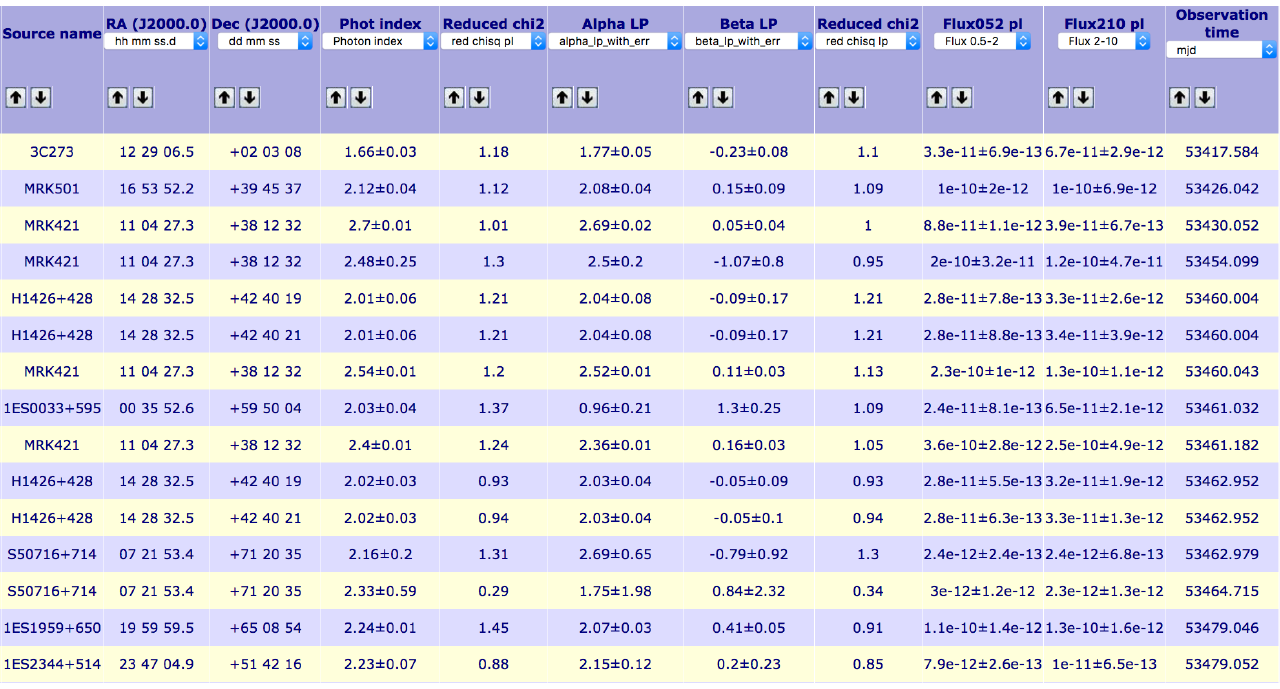}
\caption{A selection 15 lines from the on-line interactive and Virtual Observatory inter-operable table. The results of the analysis of all the spectral fits of the 65 blazars listed in table \ref{TheSample} are accessible at 
https://openuniverse.asi.it/blazars/swift.
A fits file including all the results is available at https://openuniverse.asi.it/OU4Blazars/oublazars\_swift\_spectra\_v1.0.fits.gz.
}
\label{fig:onlinetable}       
\end{figure*}

\begin{figure}
\centering
\includegraphics[width=8.5cm]{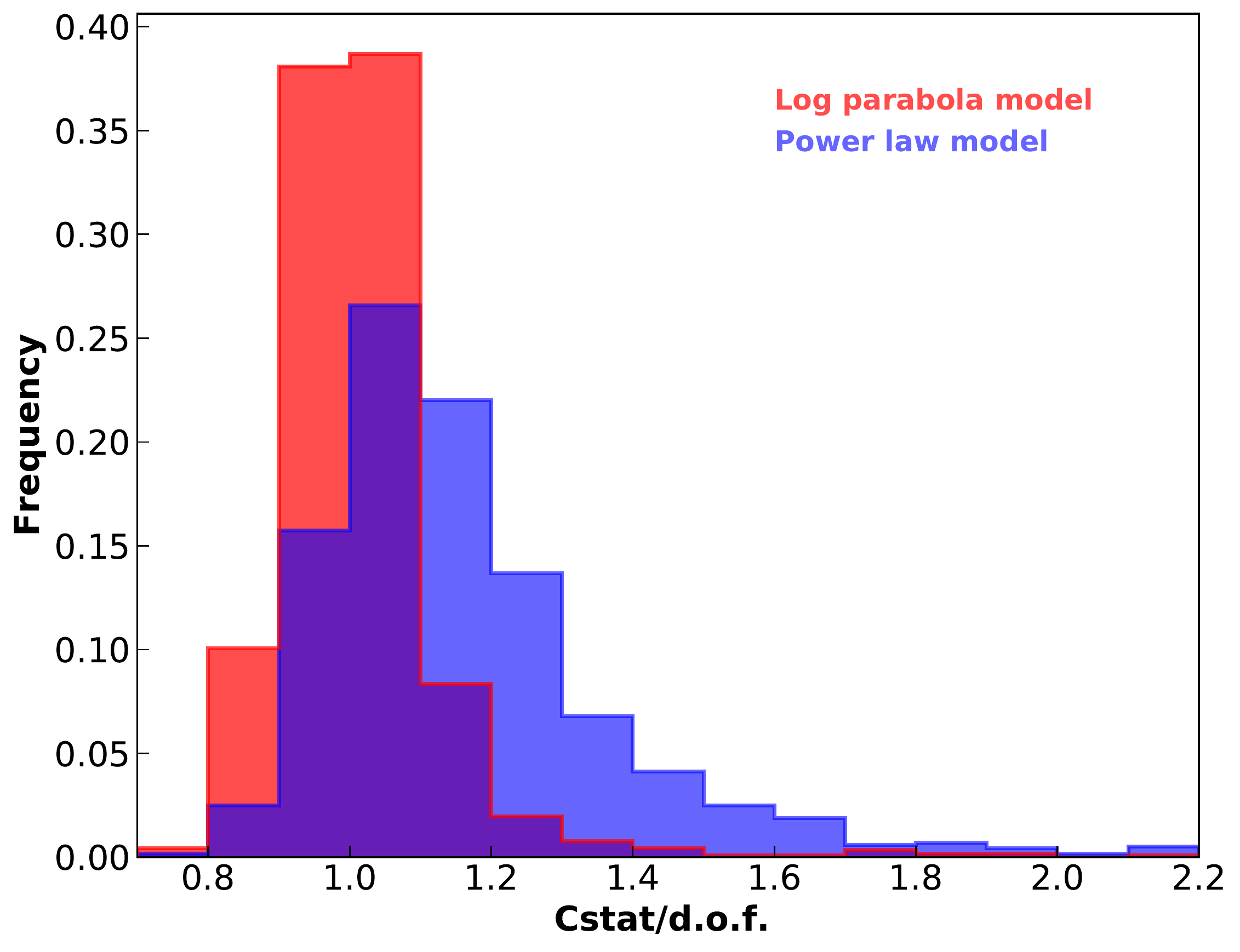}
\caption{The normalised distributions of best fit Cstat statistics divided by the number of degrees of freedom (Cstat/d.o.f.) for the power-law and log-parabola models for the case of the blazar Mrk 421. The log-parabola model gives systematically lower Cstat/d.o.f. values, implying that this model is clearly a better representation of the data for this blazar.}
\label{fig:chisqdistr}       
\end{figure} 

\begin{figure}
\centering
\includegraphics[width=8.5cm]{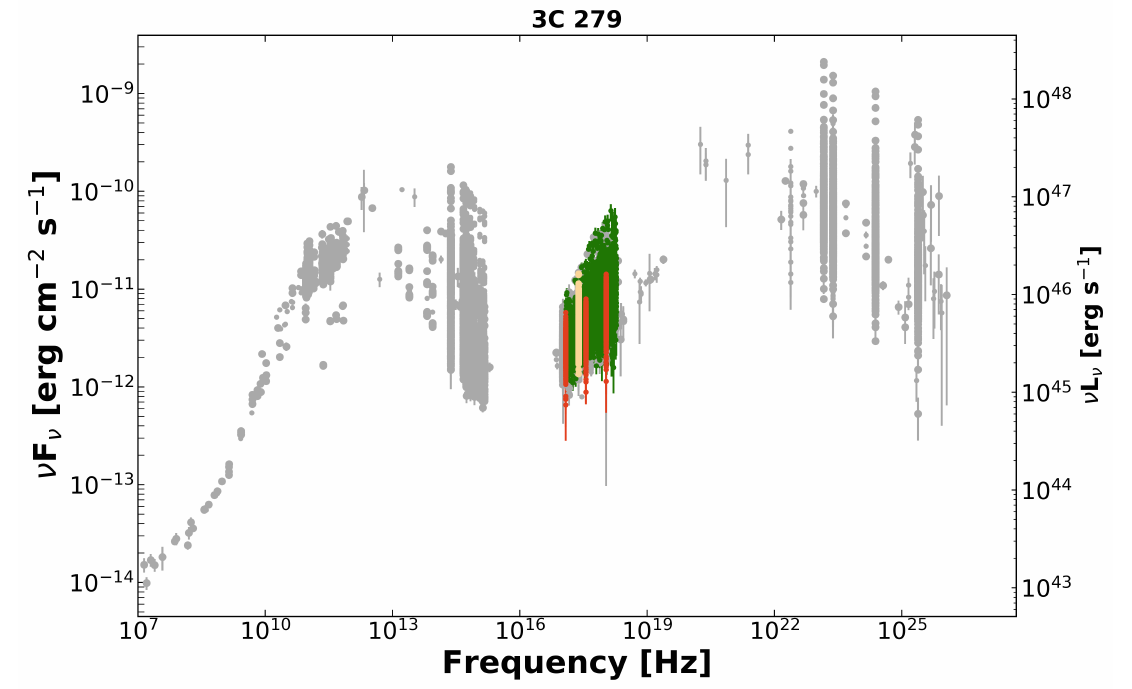}\\
\includegraphics[width=8.5cm]{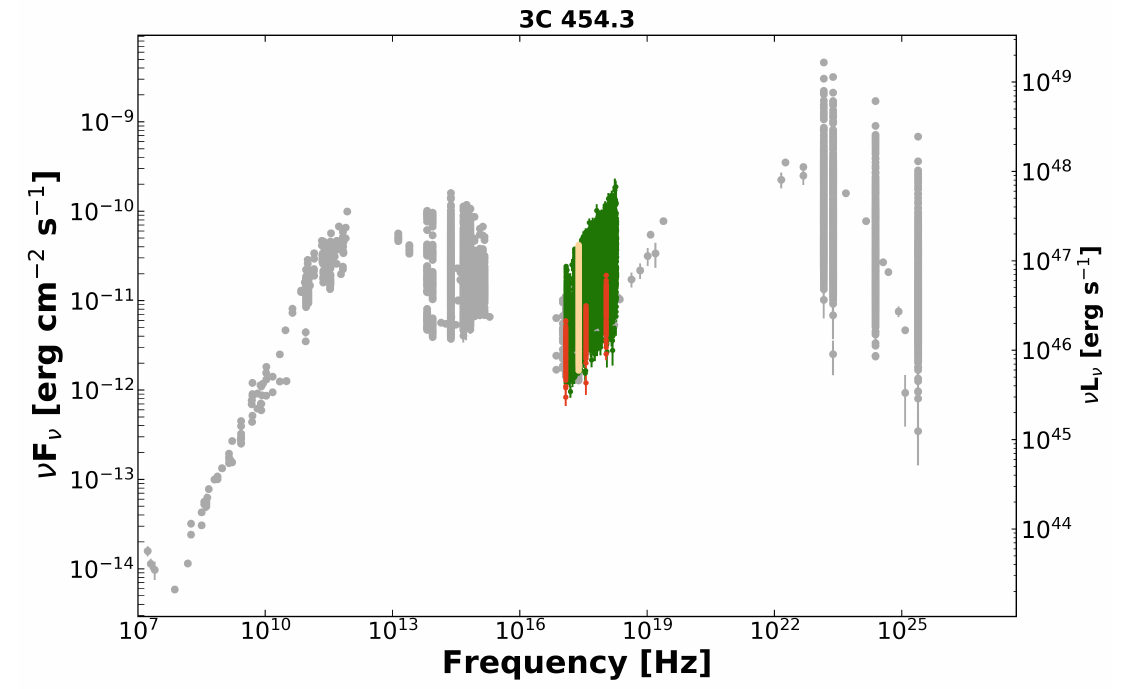} \\
\includegraphics[width=8.5cm]{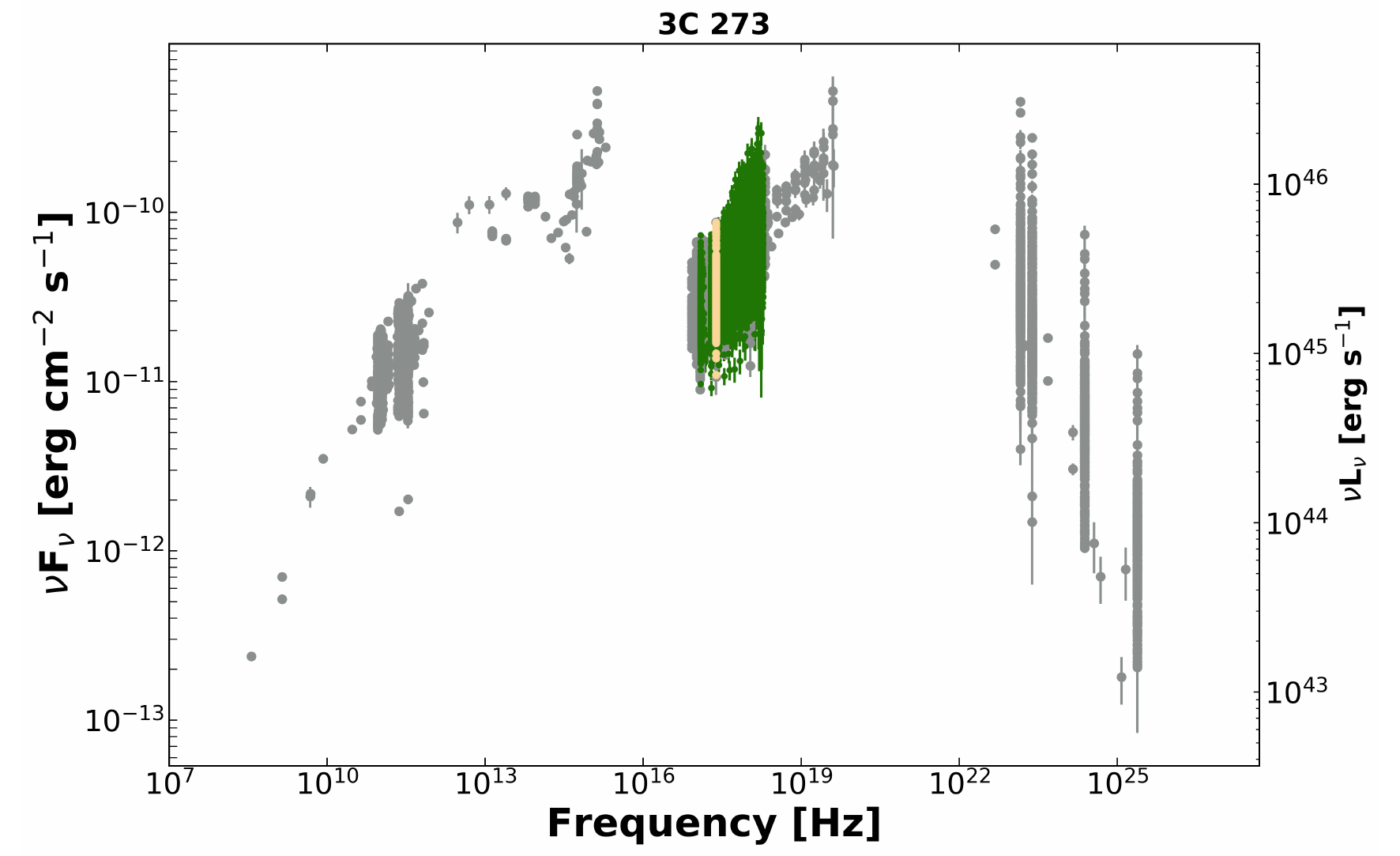}\\
\includegraphics[width=8.5cm]{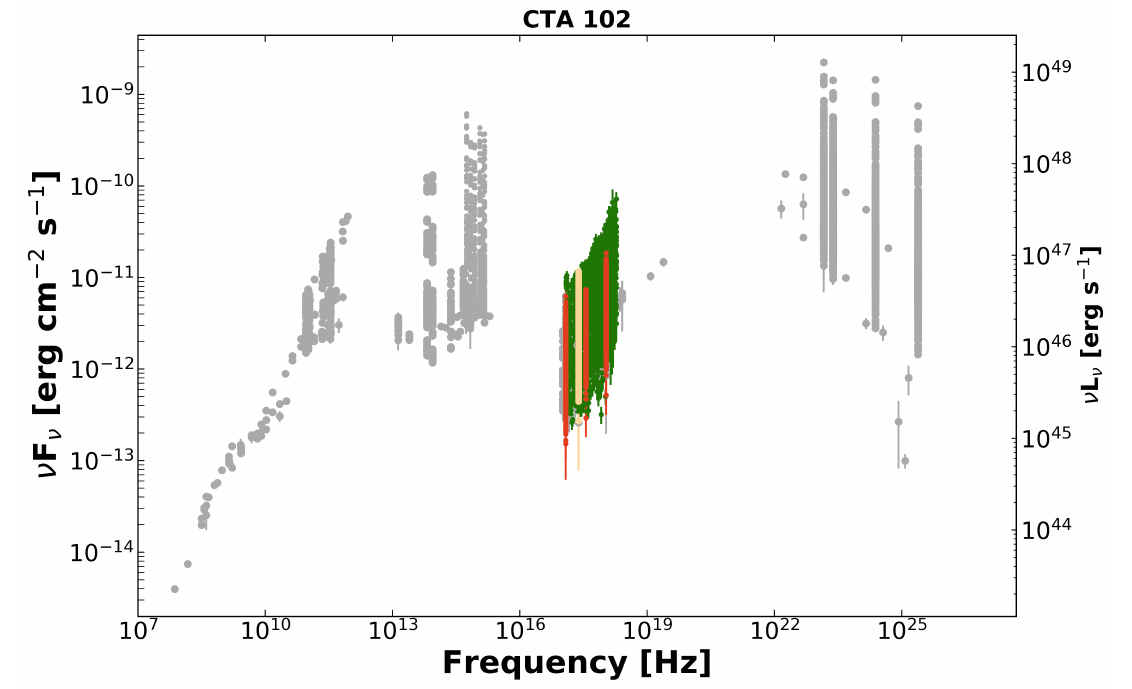}
\caption{SED of representative LBL blazars showing the XSPEC best fit spectra (WT and PC mode, green points), the results of the XIMAGE photometric analysis (only PC mode, red points), the overall 1keV light curve points (light yellow points) and archival multi-frequency data from VOU-Blazars (grey points).
}
\label{fig:SEDsLBLs}       
\end{figure} 

\begin{figure}
\centering
\includegraphics[width=8.5cm]{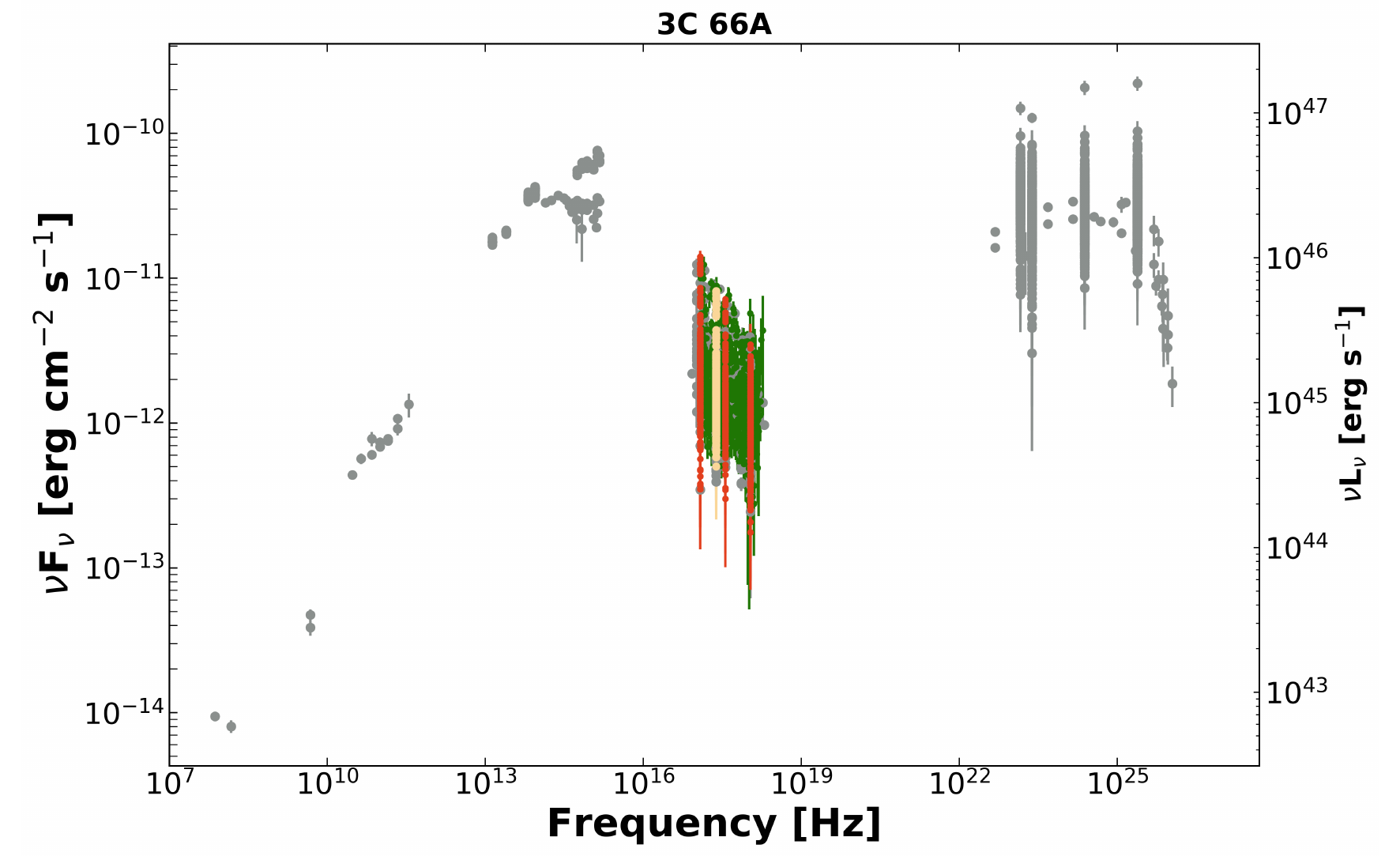}\\
\includegraphics[width=8.5cm]{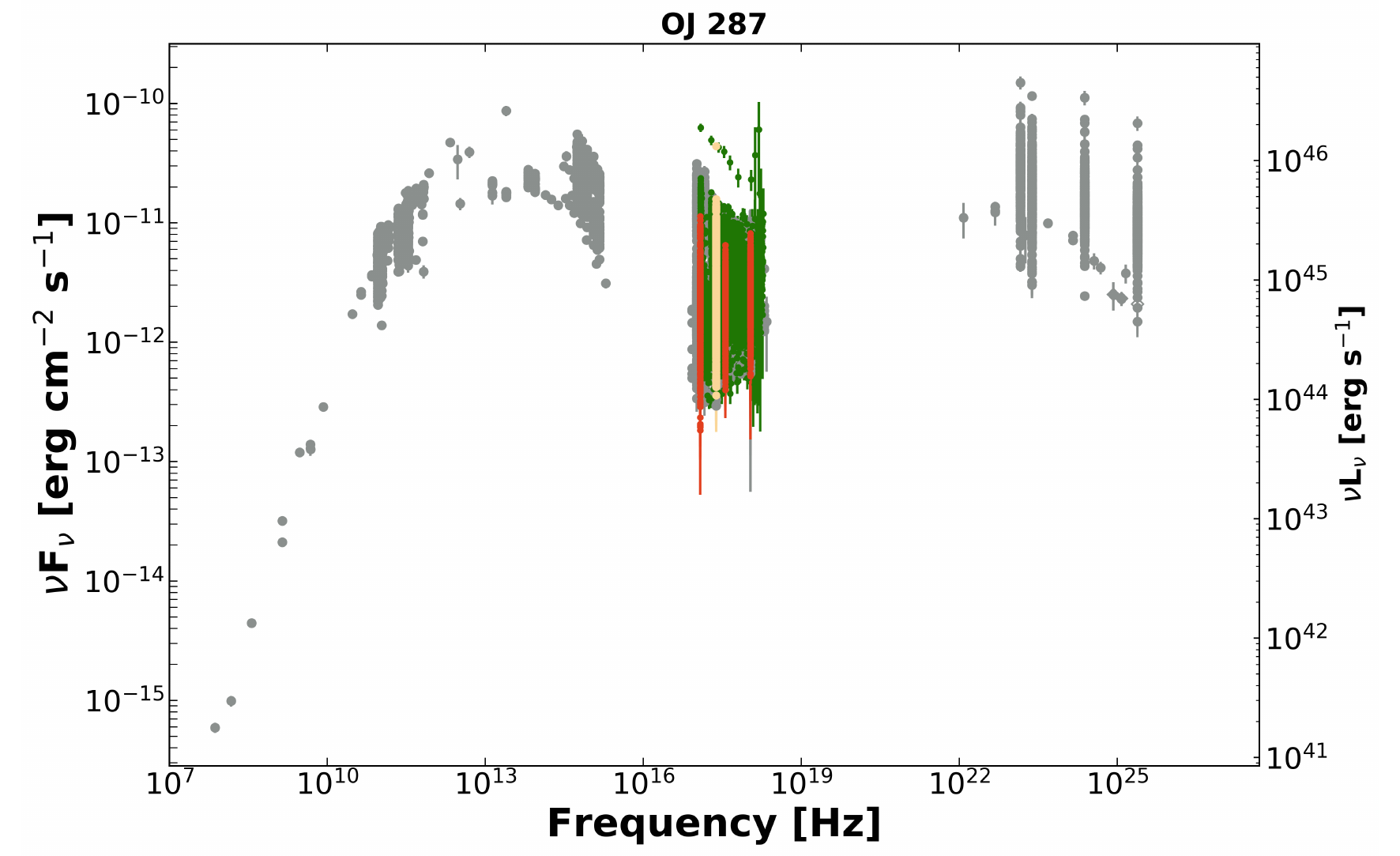}\\
\includegraphics[width=8.5cm]{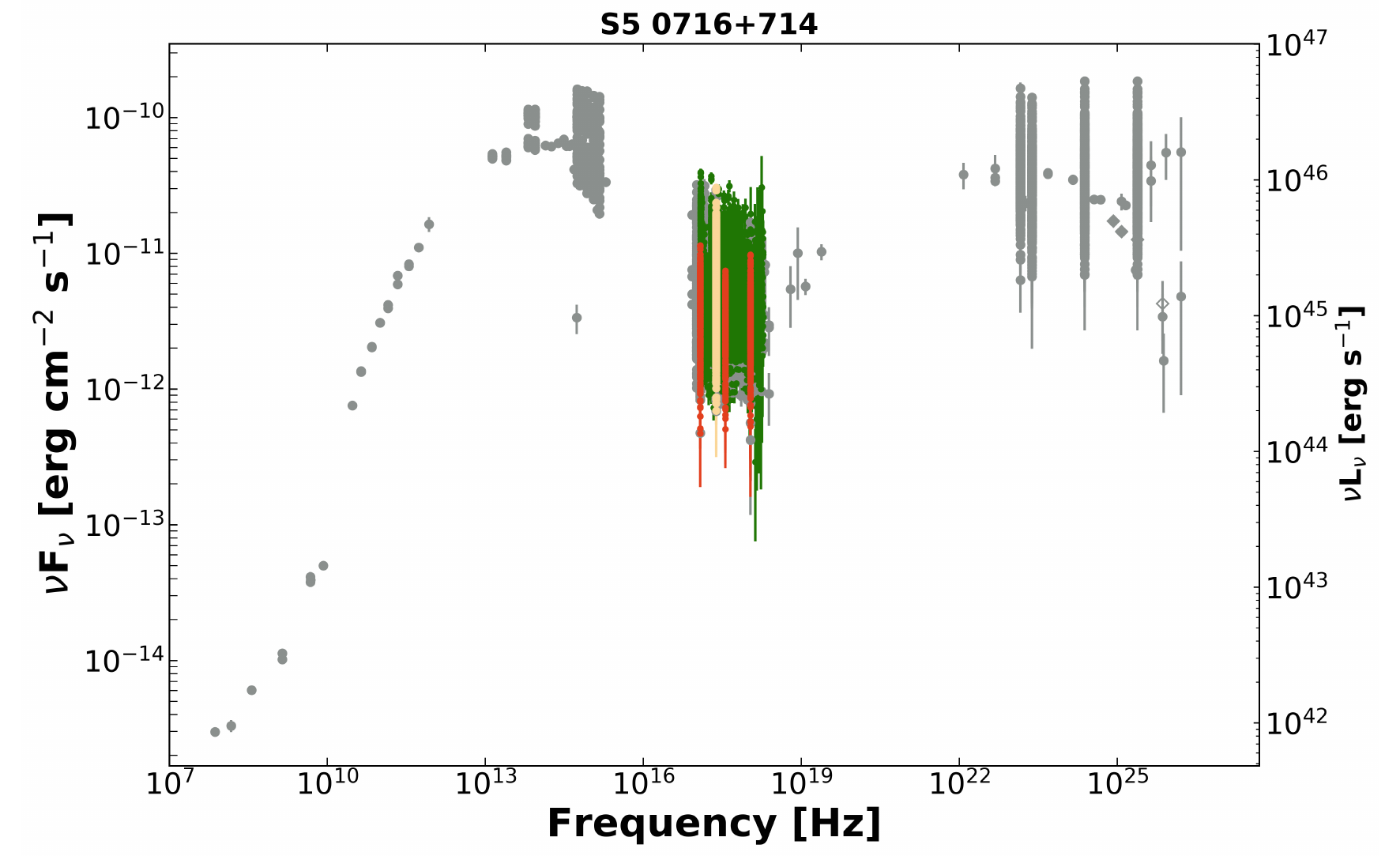}\\
\includegraphics[width=8.5cm]{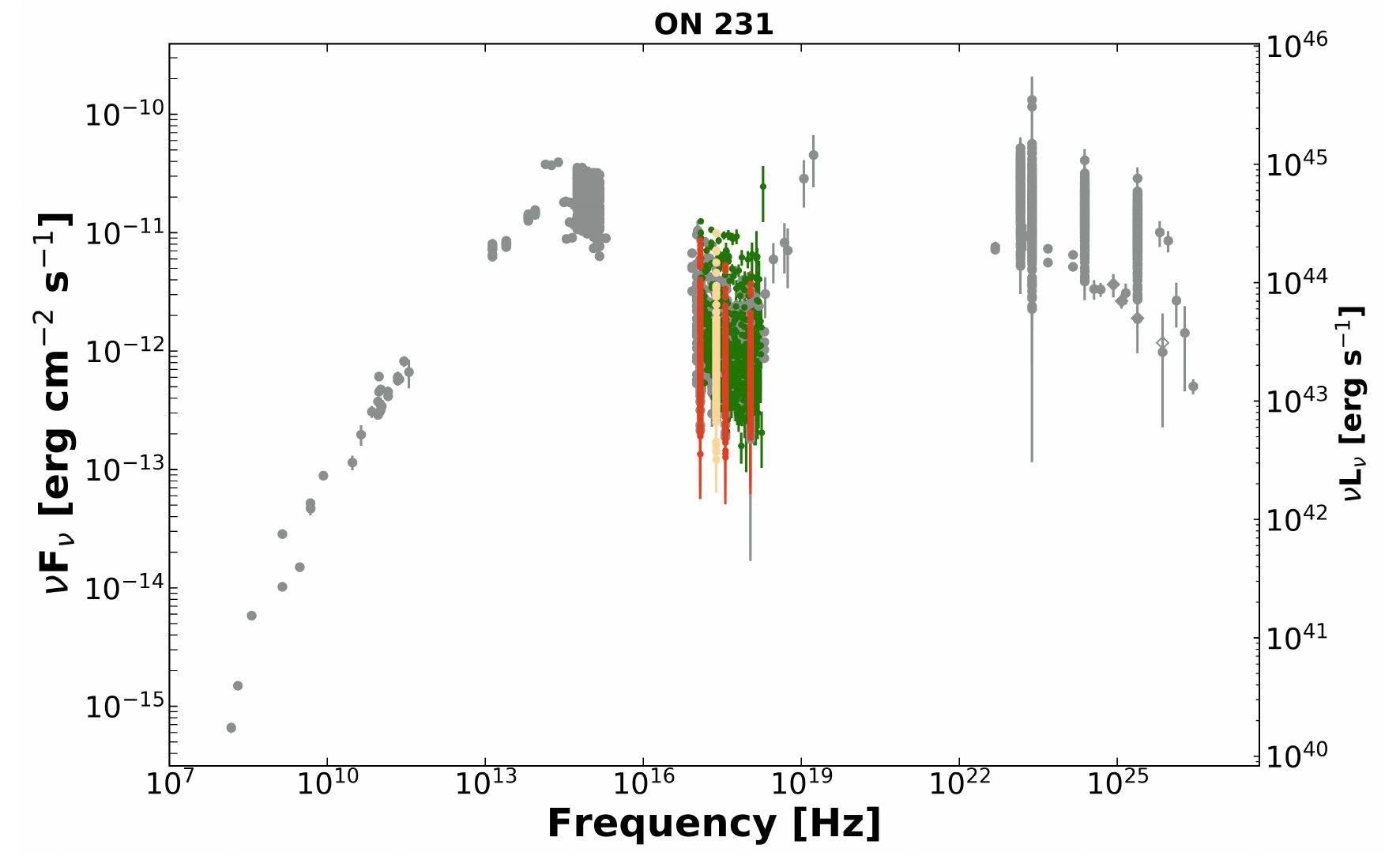}
\caption{SED of representative IBL blazars showing the XSPEC best fit spectra (WT and PC mode, green points), the results of the XIMAGE photometric analysis (only PC mode, red points), the overall 1keV light curve points (light yellow points) and archival multi-frequency data from VOU-Blazars (grey points).}
\label{fig:SEDsIBLs}       
\end{figure} 

\begin{figure}
\centering
\includegraphics[width=8.5cm]{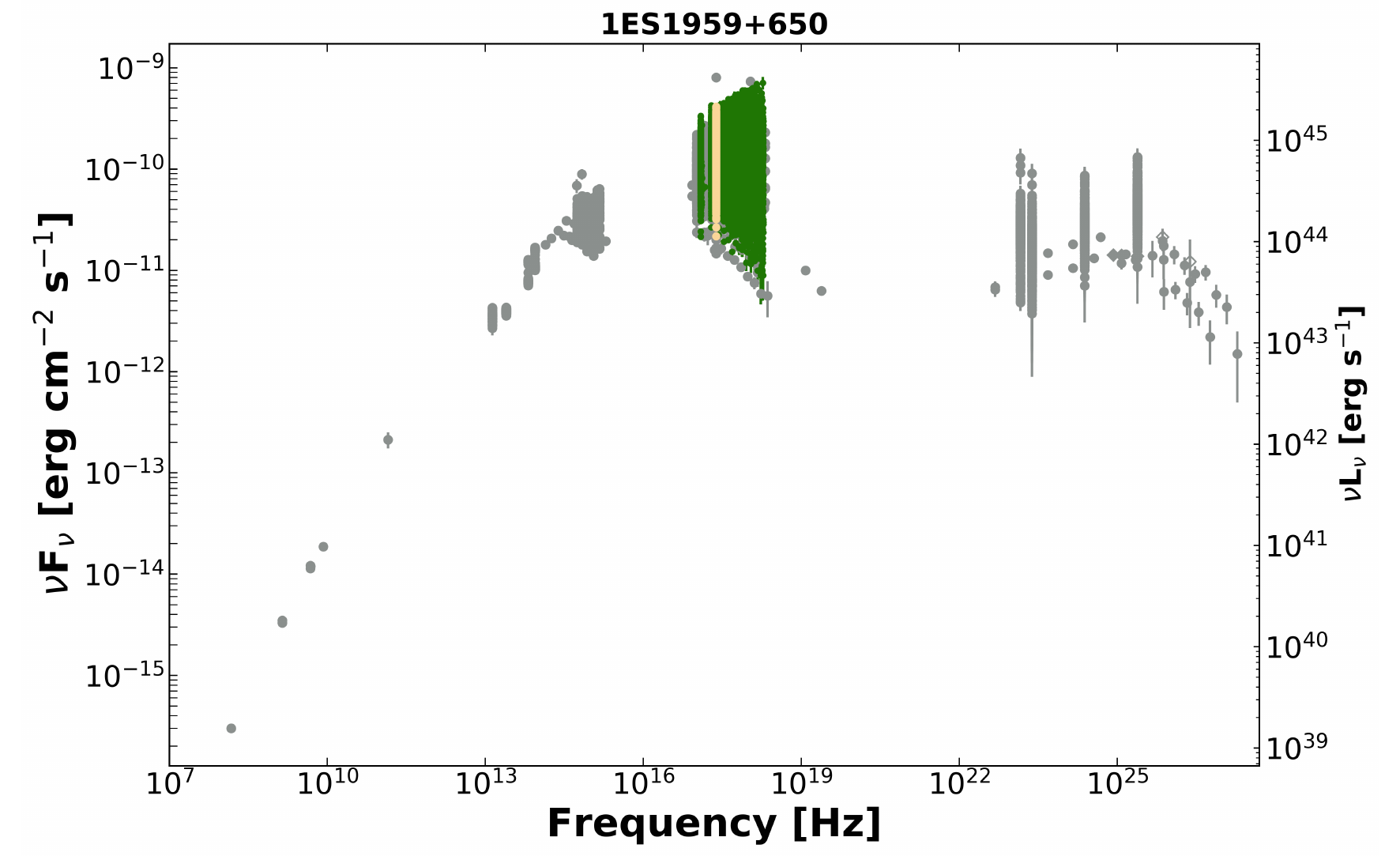}\\
\includegraphics[width=8.5cm]{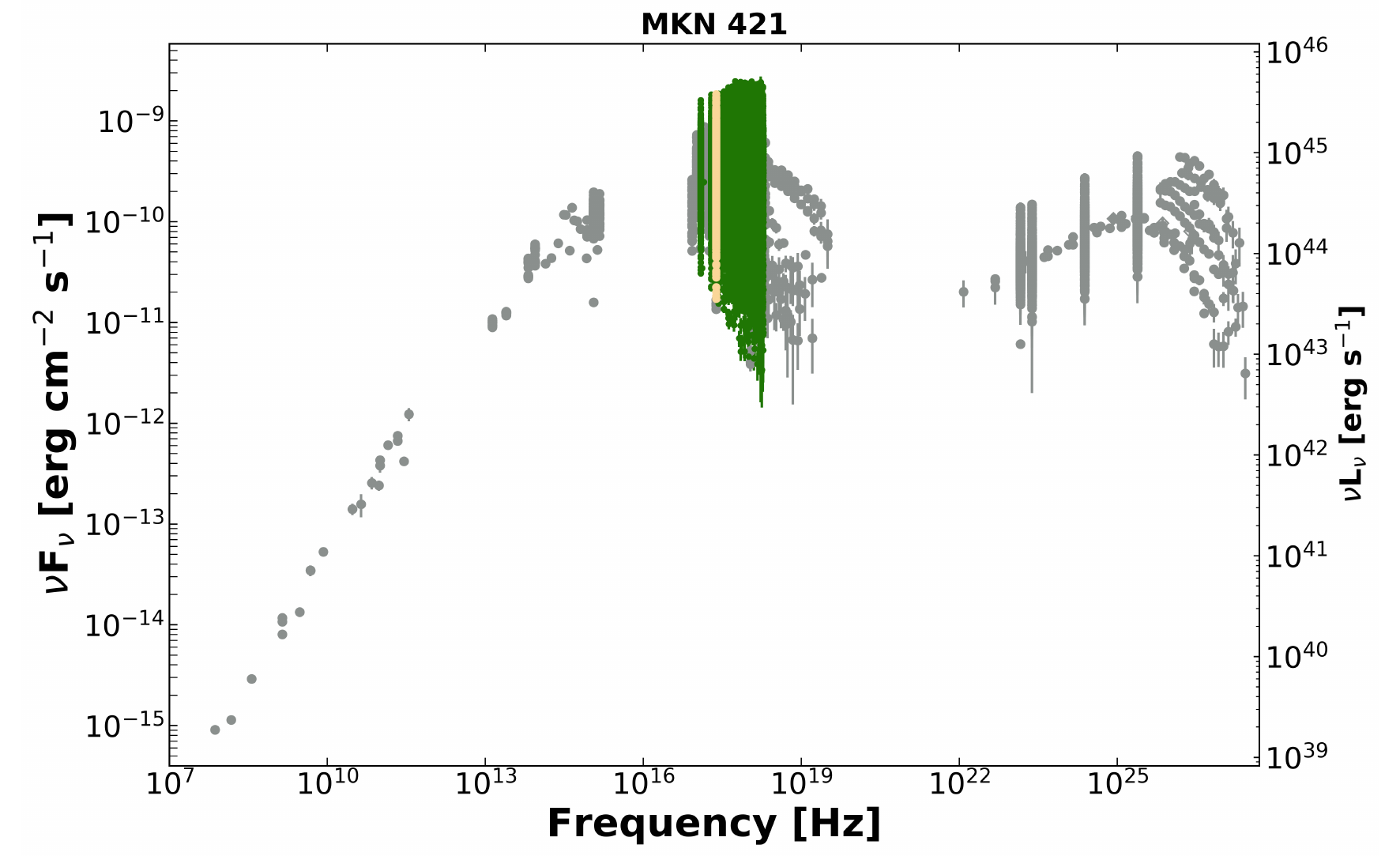} \\
\includegraphics[width=8.5cm]{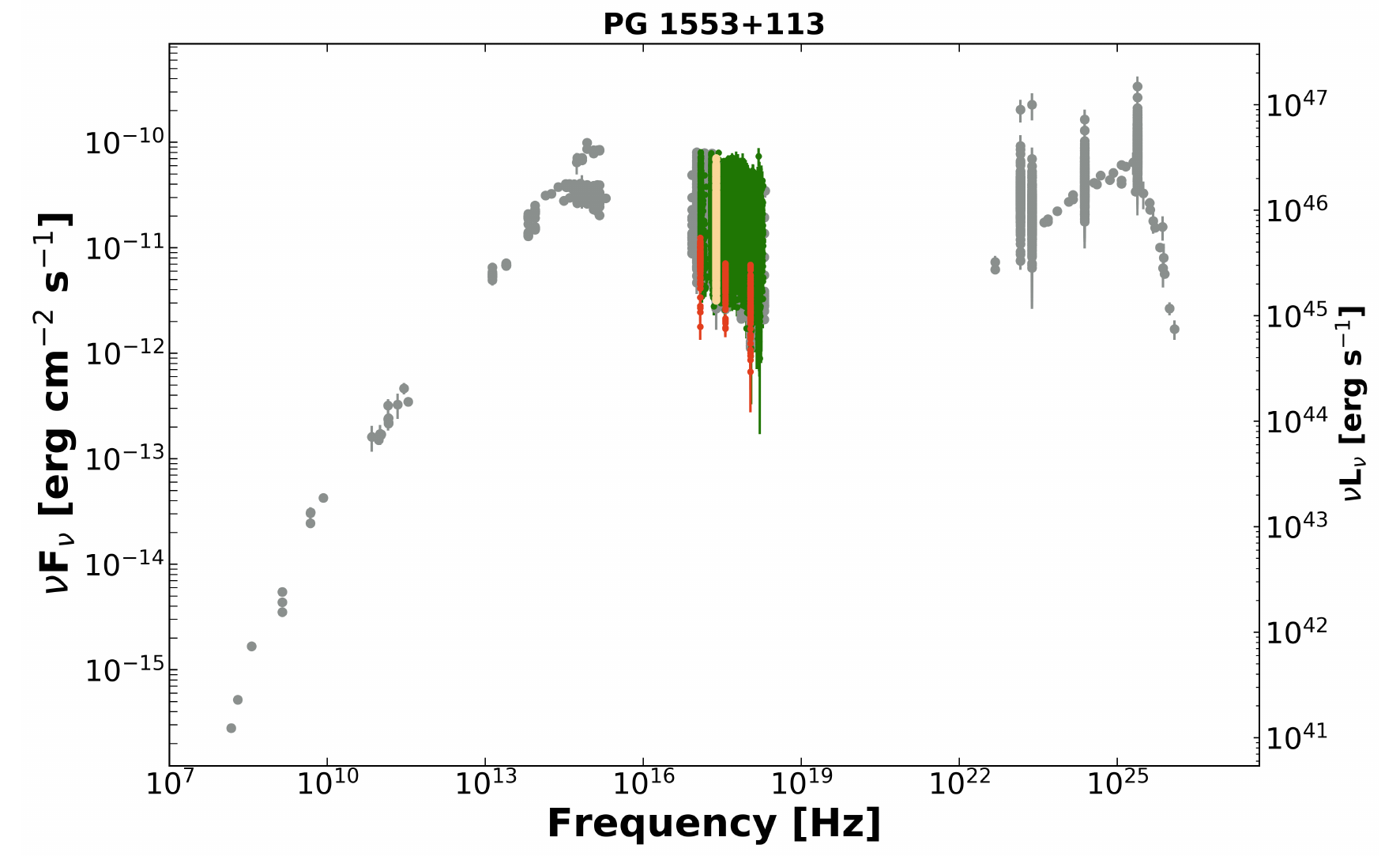} \\
\includegraphics[width=8.5cm]{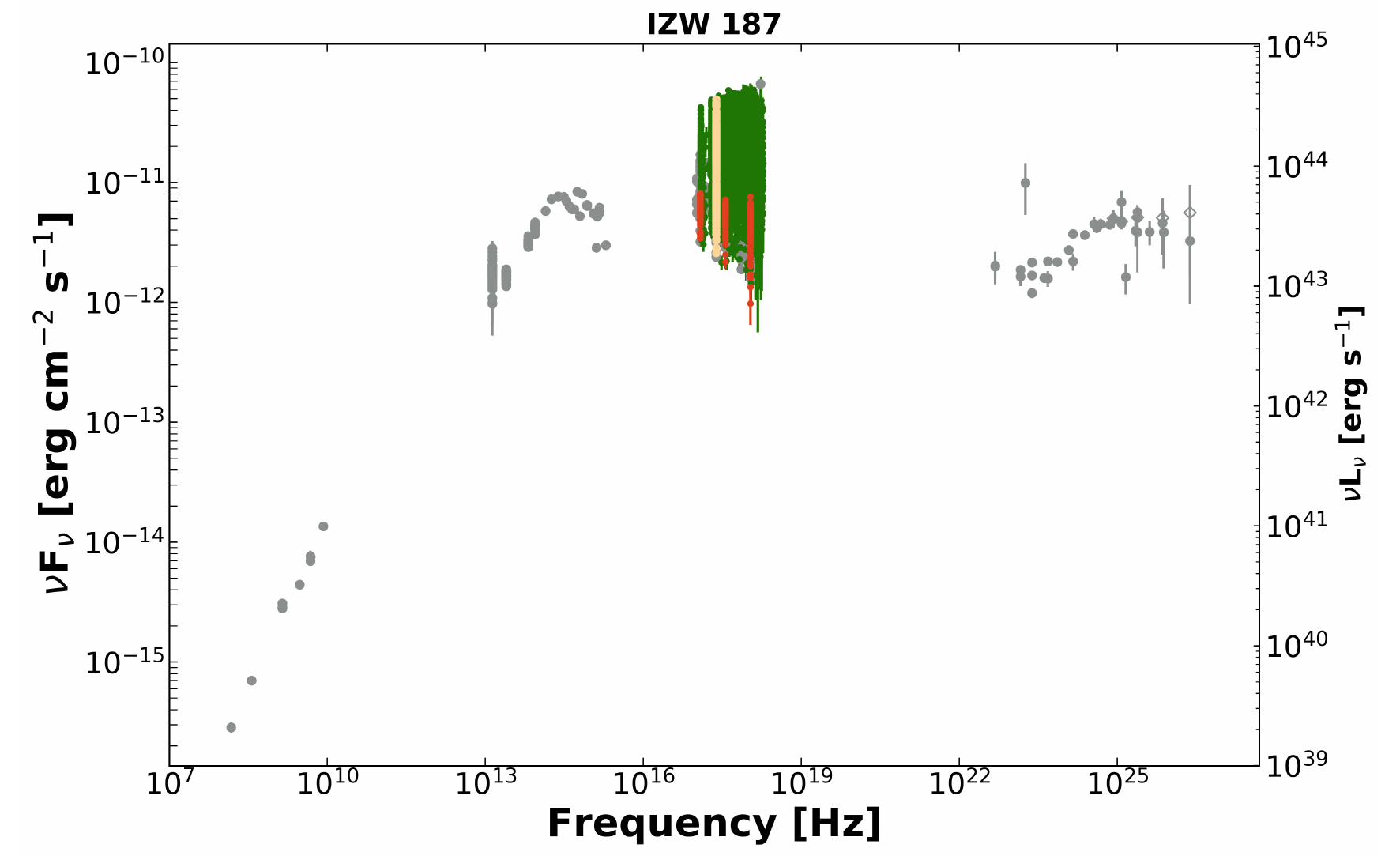} 
\caption{SED of representative HBL blazars showing the XSPEC best fit spectra (WT and PC mode, green points), the results of the XIMAGE photometric analysis (only PC mode, red points), the overall 1keV light curve points (light yellow points) and archival multi-frequency data from VOU-Blazars (grey points).}
\label{fig:SEDsHBLs}       
\end{figure} 

\begin{figure}
\centering
\includegraphics[width=8.5cm]{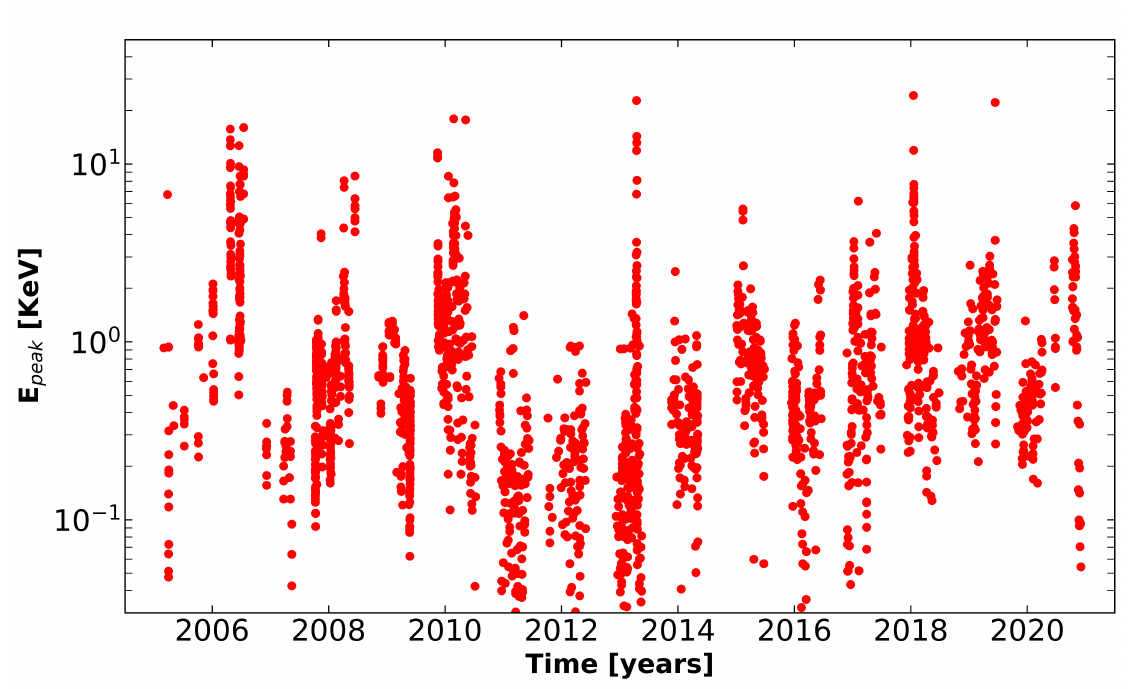}
\caption{The remarkable changes of the synchrotron peak energy in the SED of Mrk 421 over a period of 16 years of Swift monitoring of this object.}
\label{fig:mkn421-epeak}       
\end{figure}

\begin{figure}
\centering
\includegraphics[width=8.5cm]{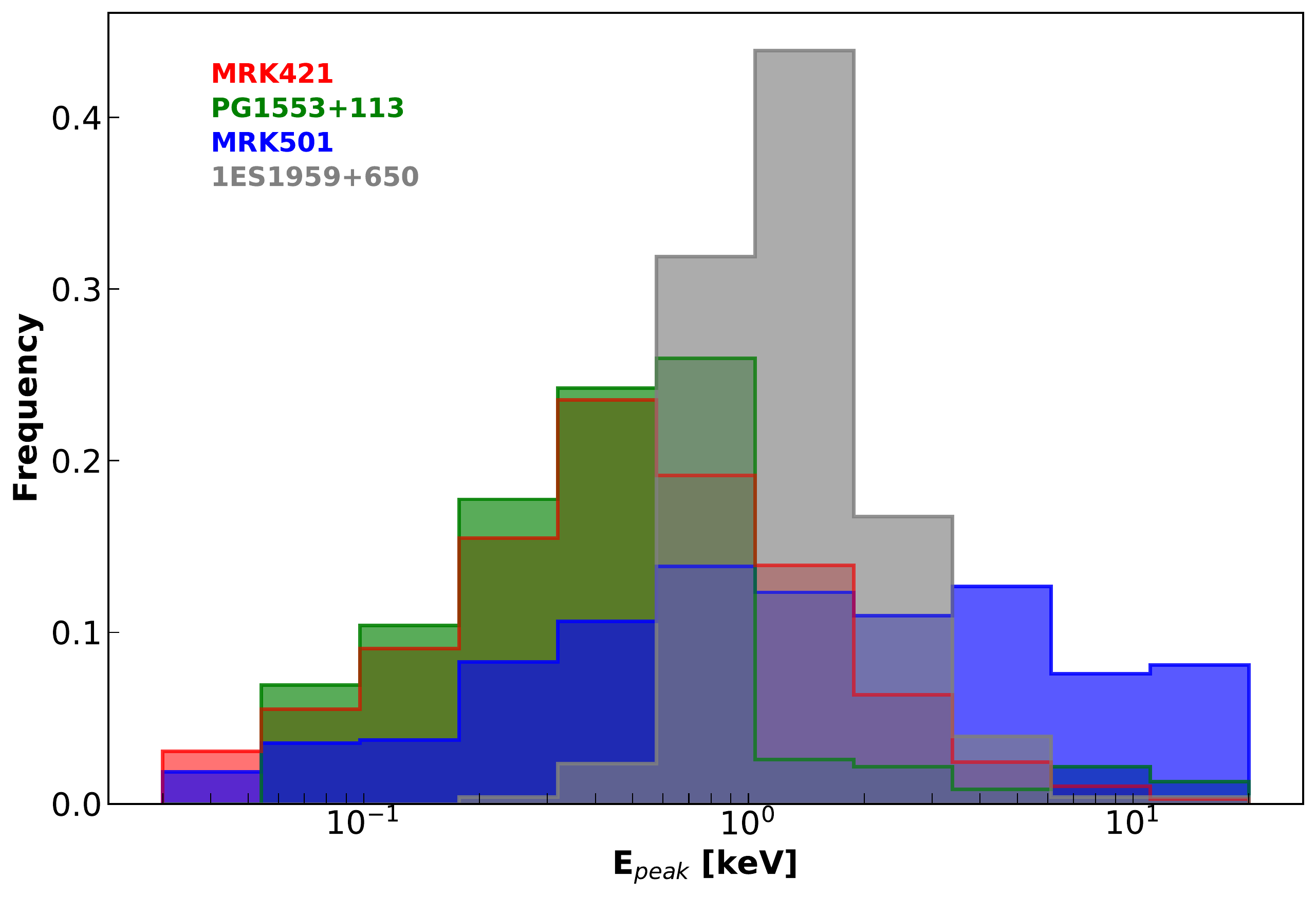}
\caption{The distribution of the synchrotron peak energy observed in bright HBL blazars as examples representing the class. Large \nupeak\, variations are commonly observed. Most of the other HBL sources in our sample show similar \nupeak\, variations.}
\label{fig:nupeak_distr}       
\end{figure}

\begin{figure}
\includegraphics[width=8.5cm]{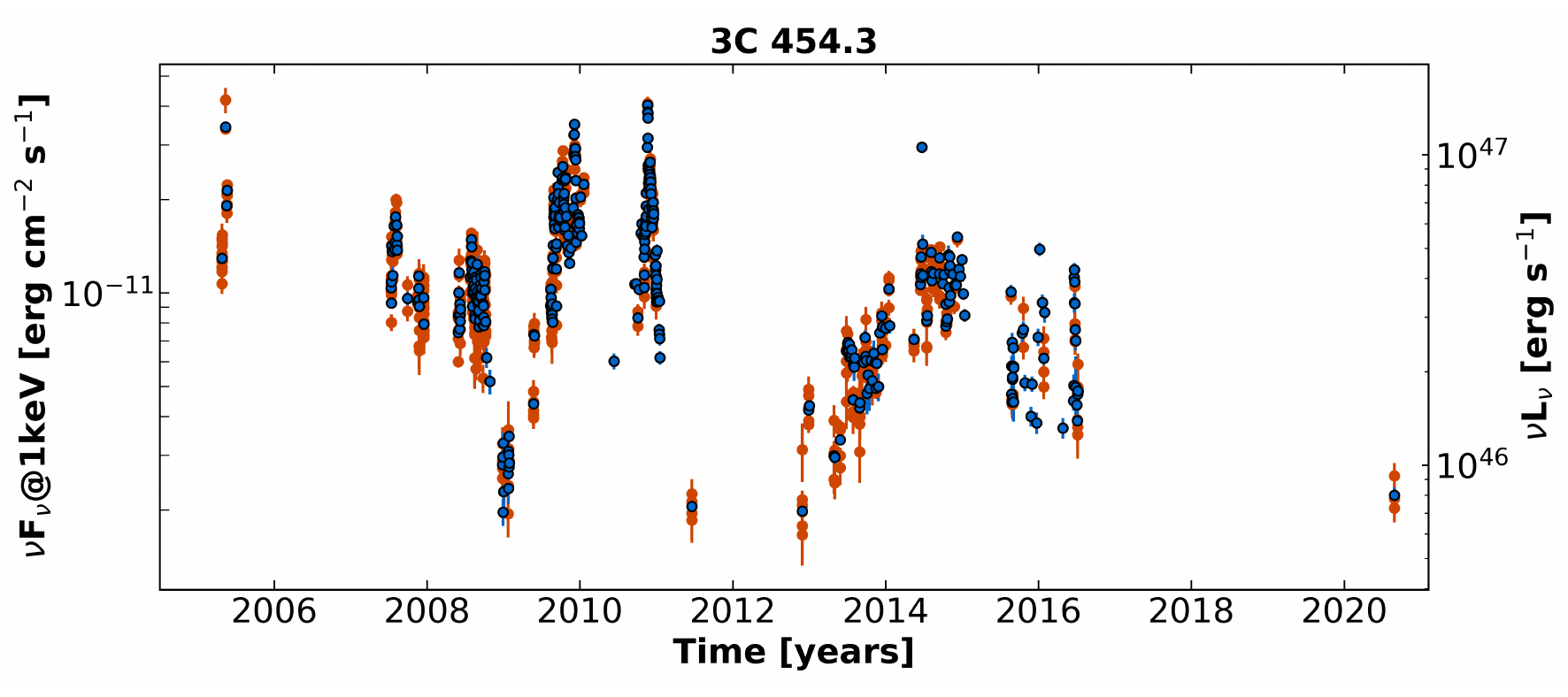} \\
\includegraphics[width=8.5cm]{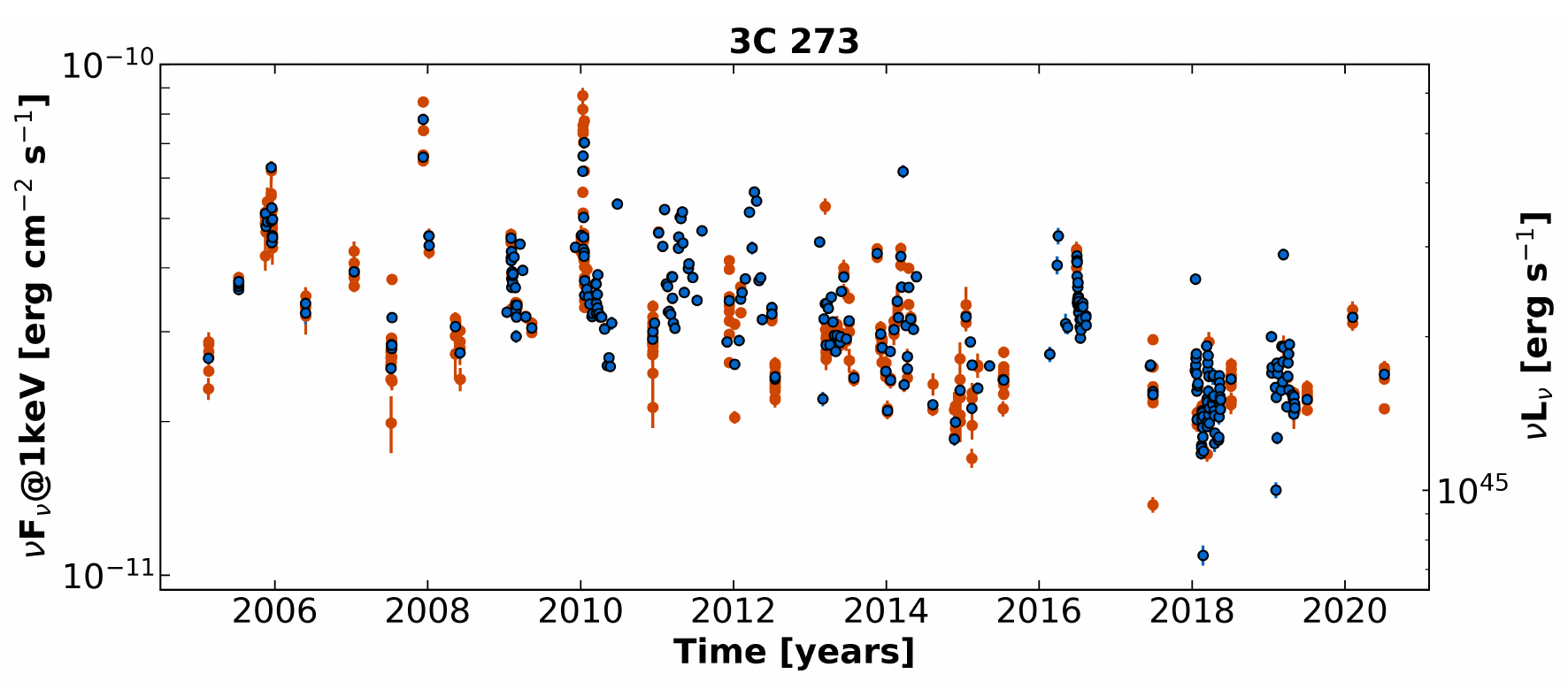} \\
\includegraphics[width=8.5cm]{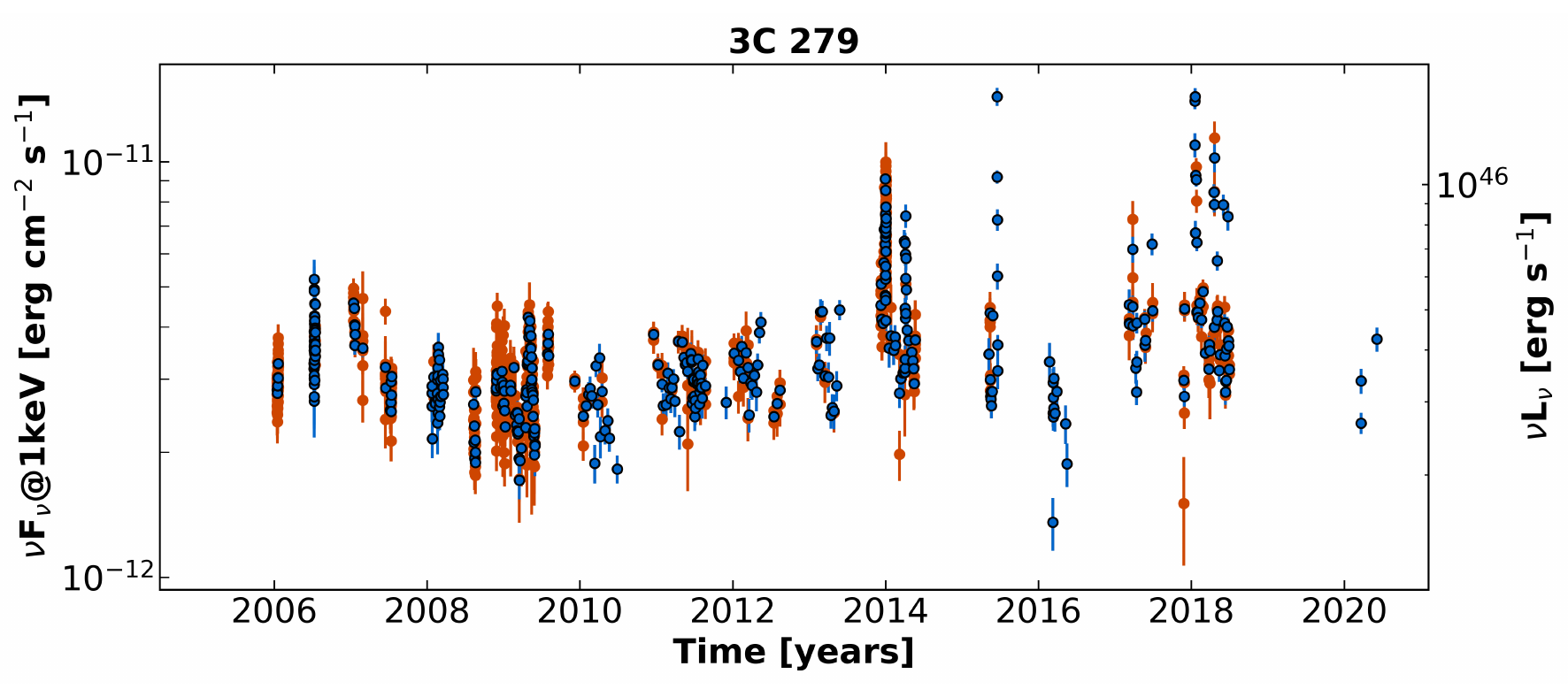} \\
\includegraphics[width=8.5cm]{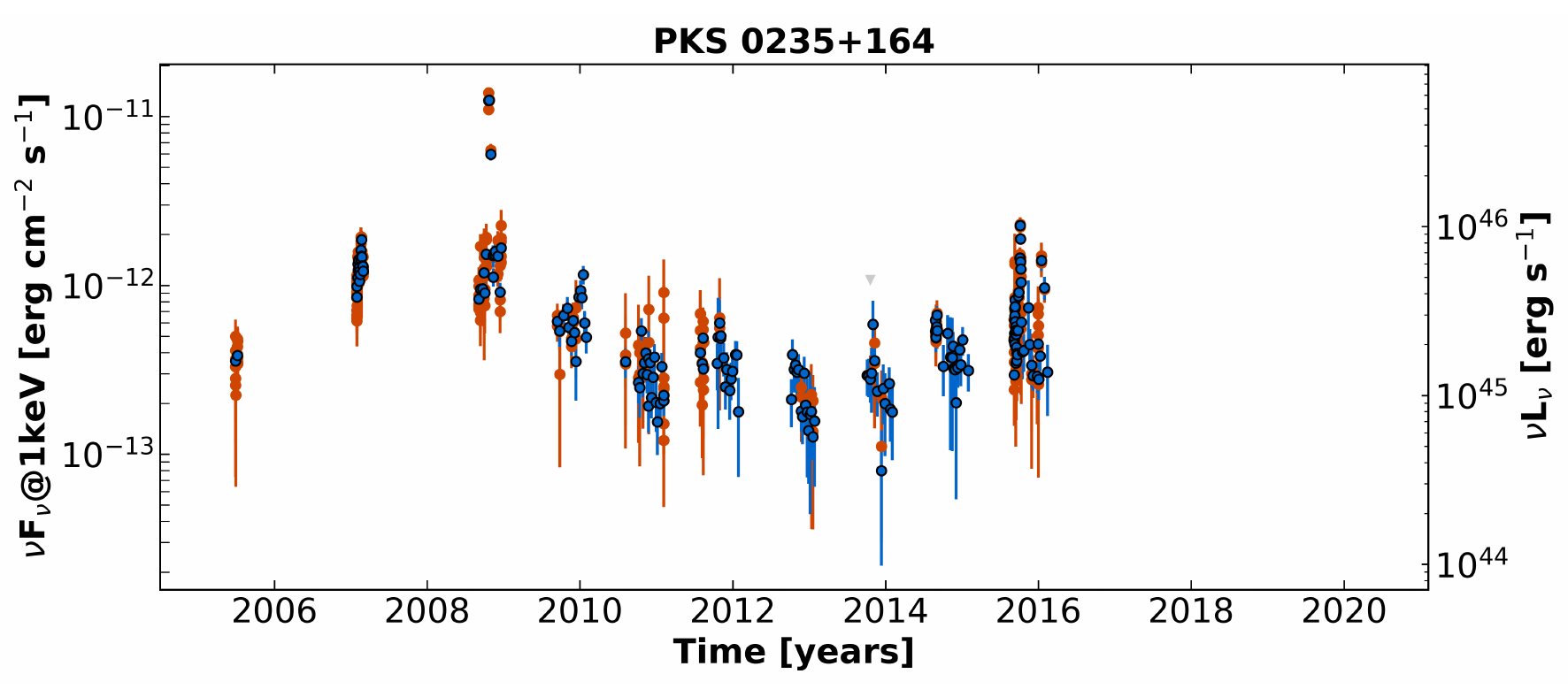}\\
\includegraphics[width=8.5cm]{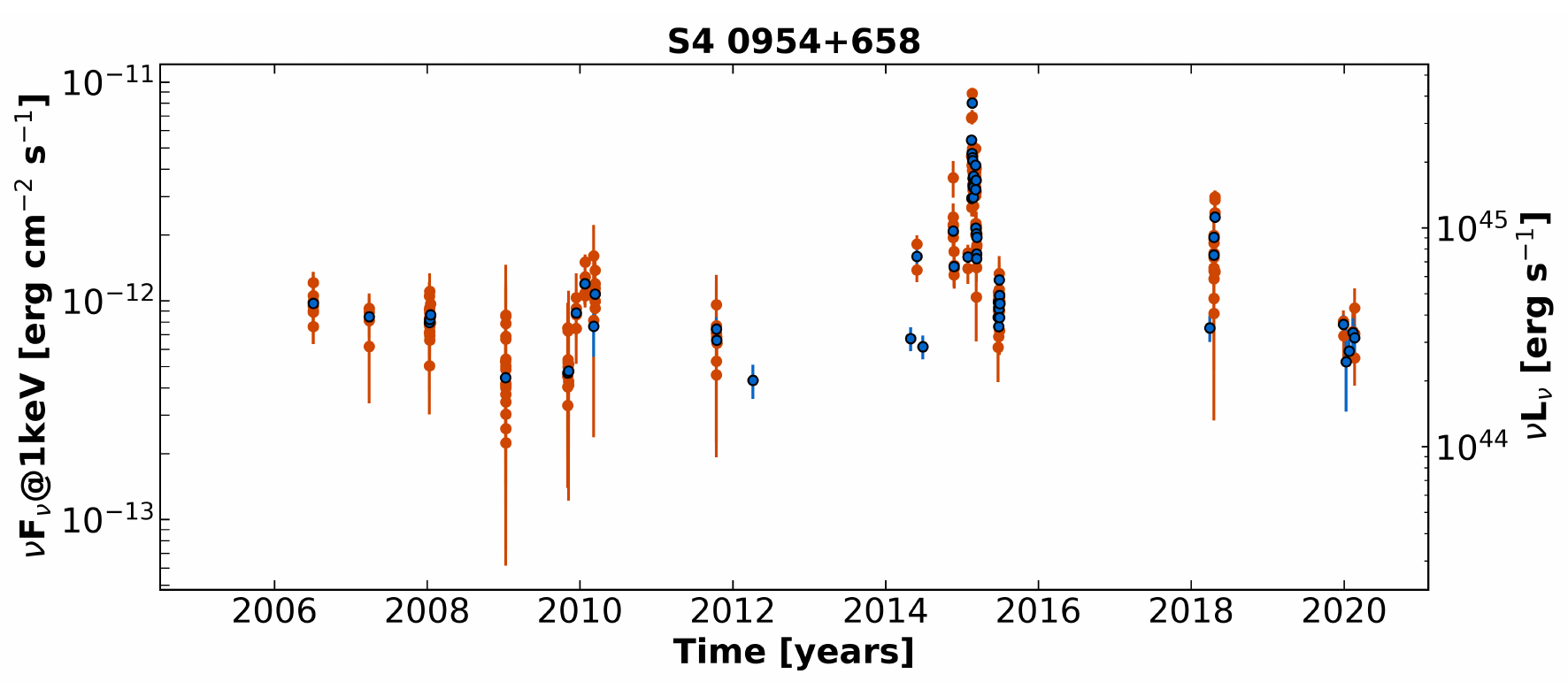}
\caption{The 1 keV light curves of LBL blazars: 3C453.4, 3C273, 3C279, PKS0235+164 and S4 0954+658}
\label{fig:LCLBLs}       
\end{figure}

\begin{figure}
\includegraphics[width=8.5cm]{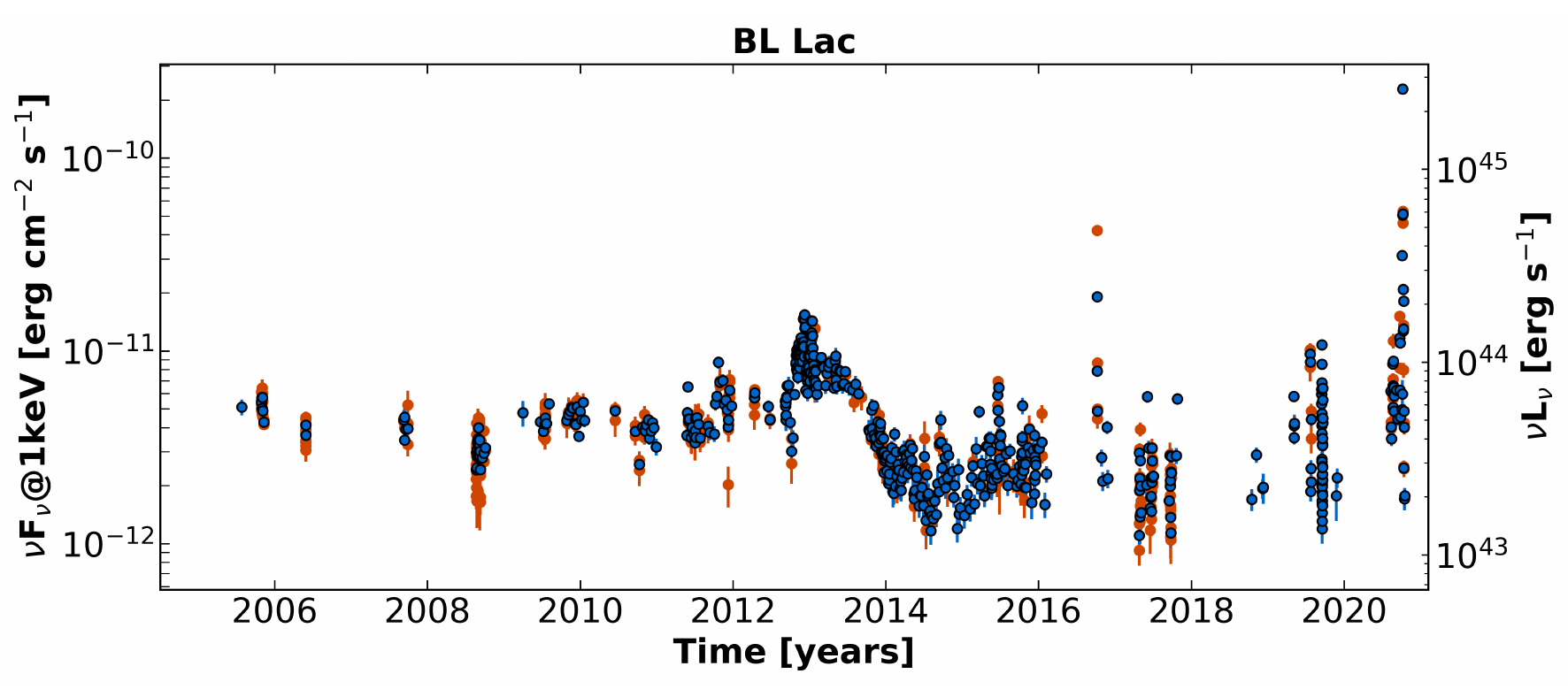}\\
\includegraphics[width=8.5cm]{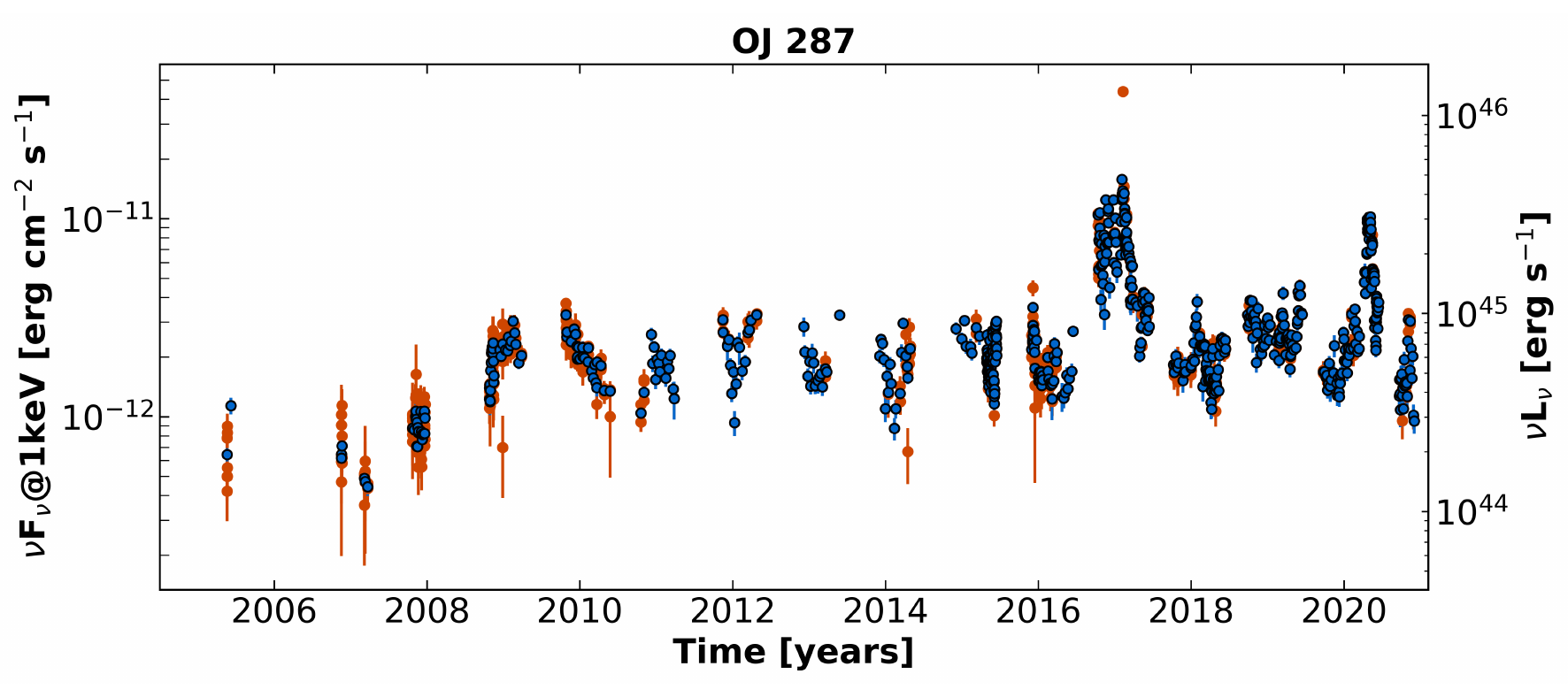}\\
\includegraphics[width=8.5cm]{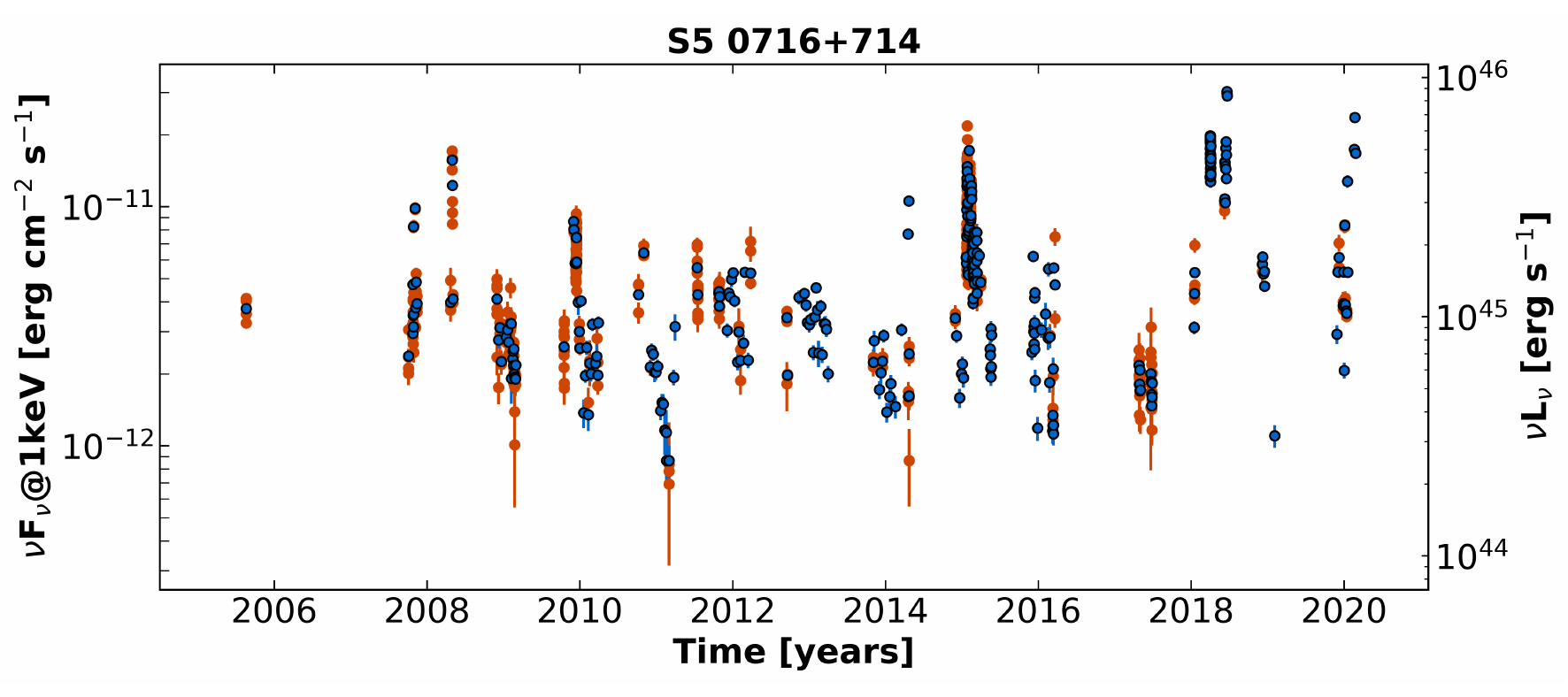}\\
\includegraphics[width=8.5cm]{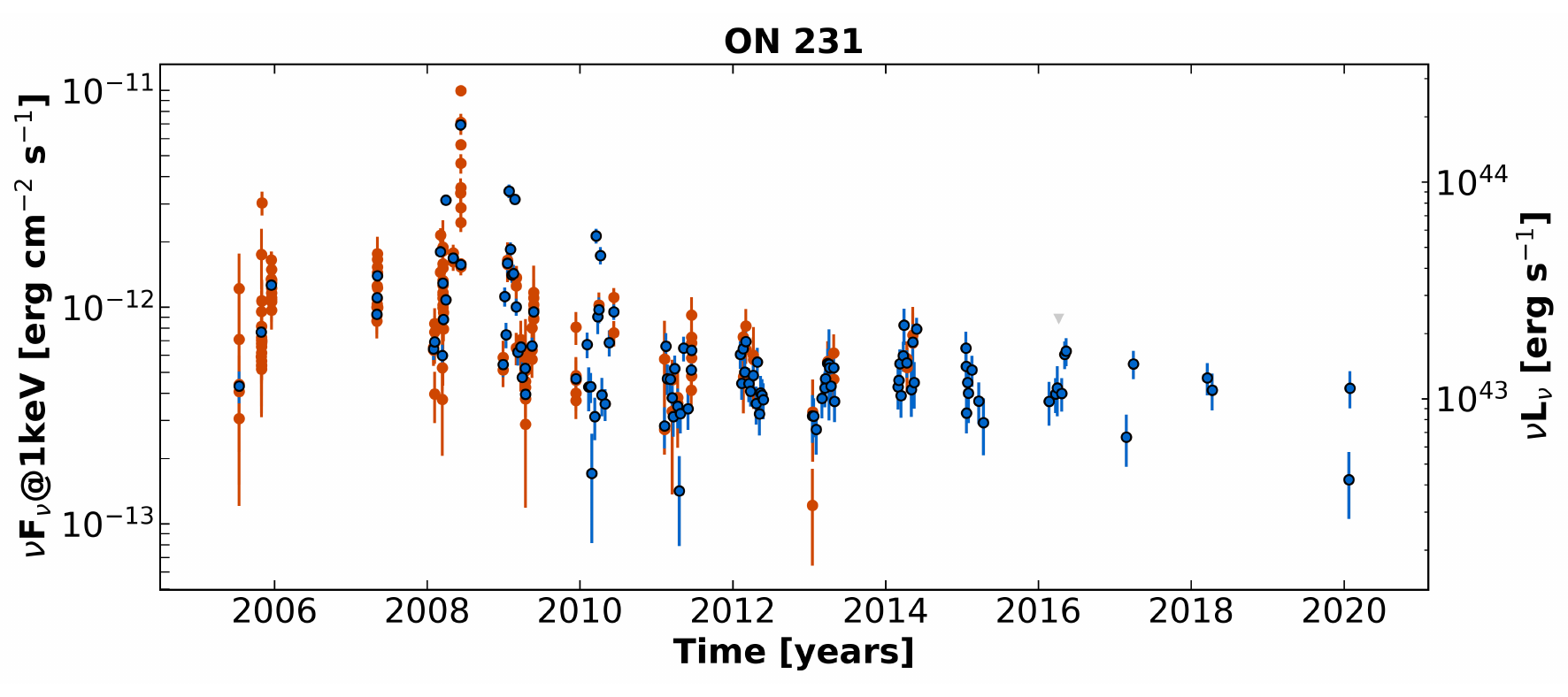} \\
\includegraphics[width=8.5cm]{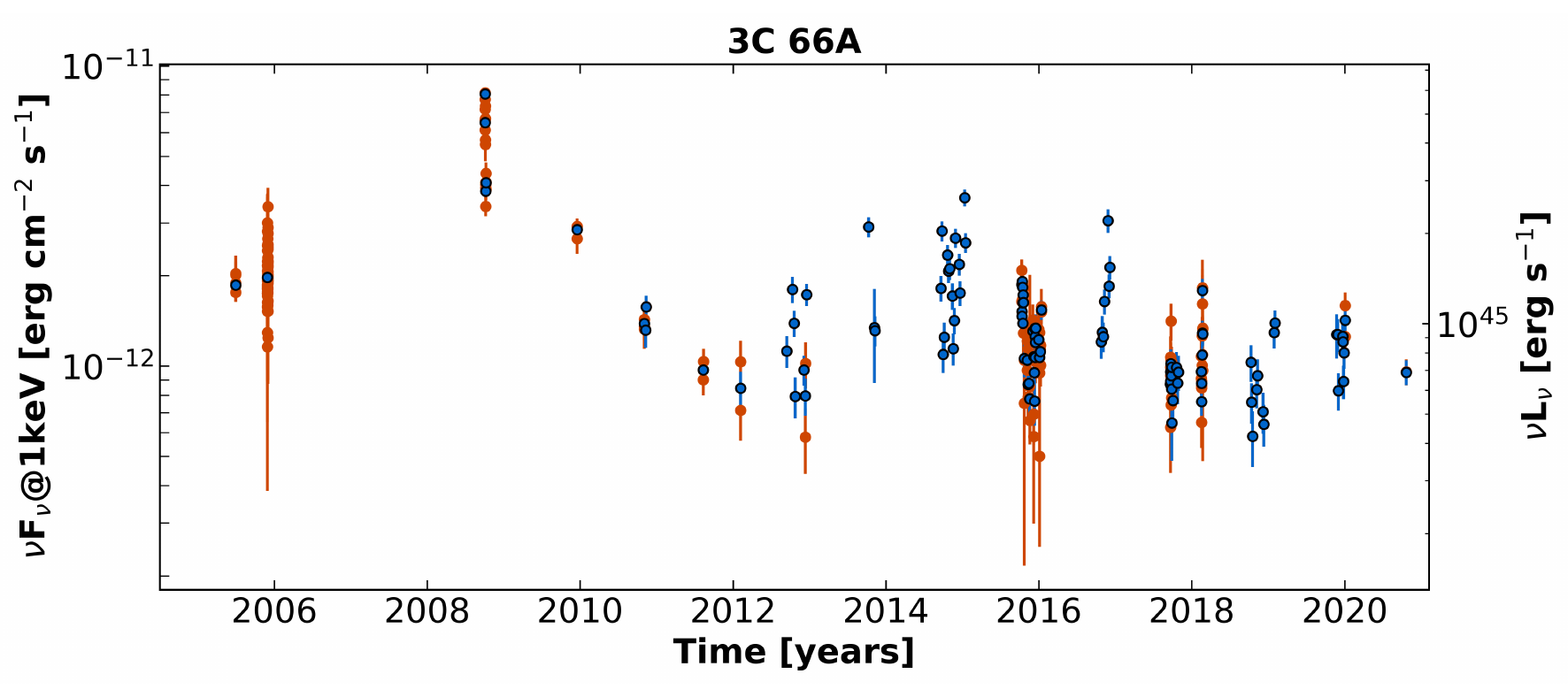}
\caption{The 1 keV light curves of IBL blazars: BL Lacertae, OJ 287, S5 0716+714, ON 231, and 3C 66A}
\label{fig:LCIBLs}       
\end{figure}

\begin{figure}
\includegraphics[width=8.5cm]{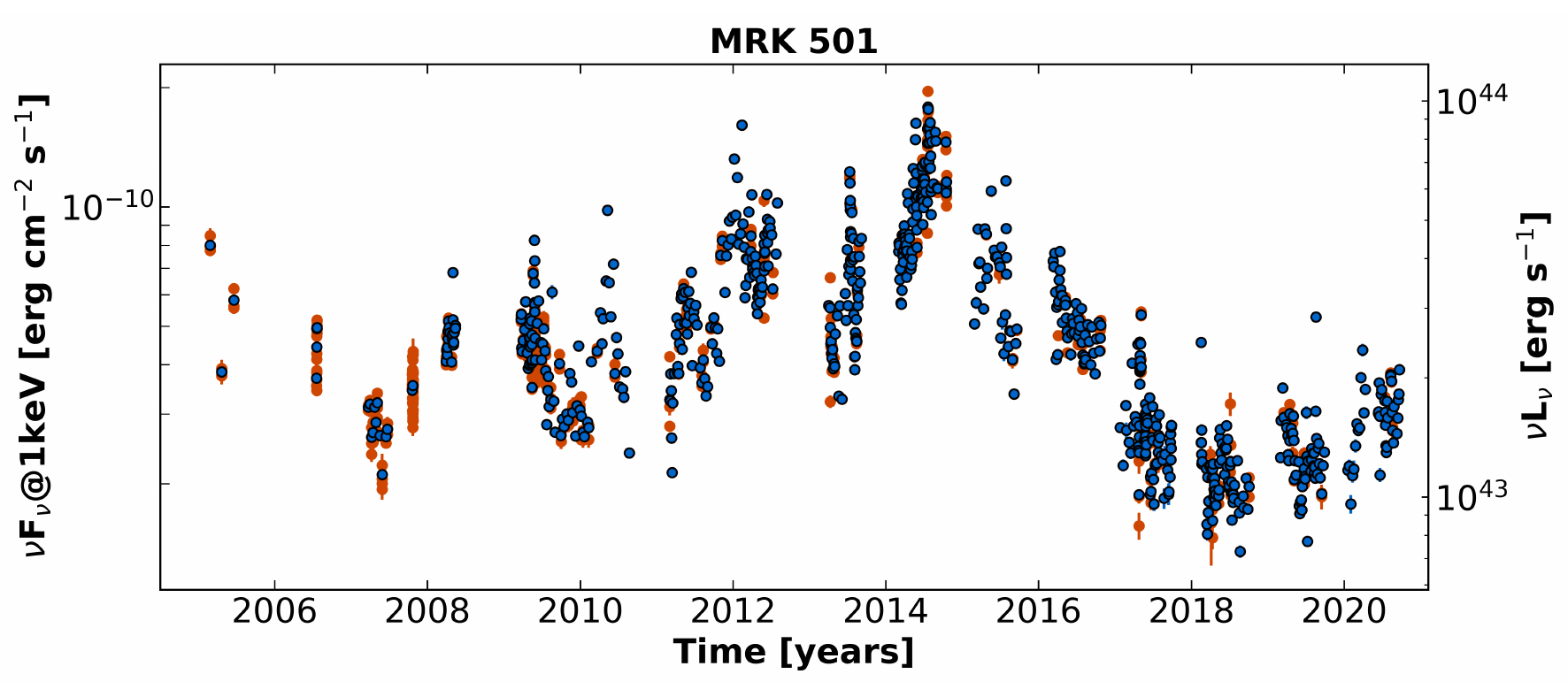}\\
\includegraphics[width=8.5cm]{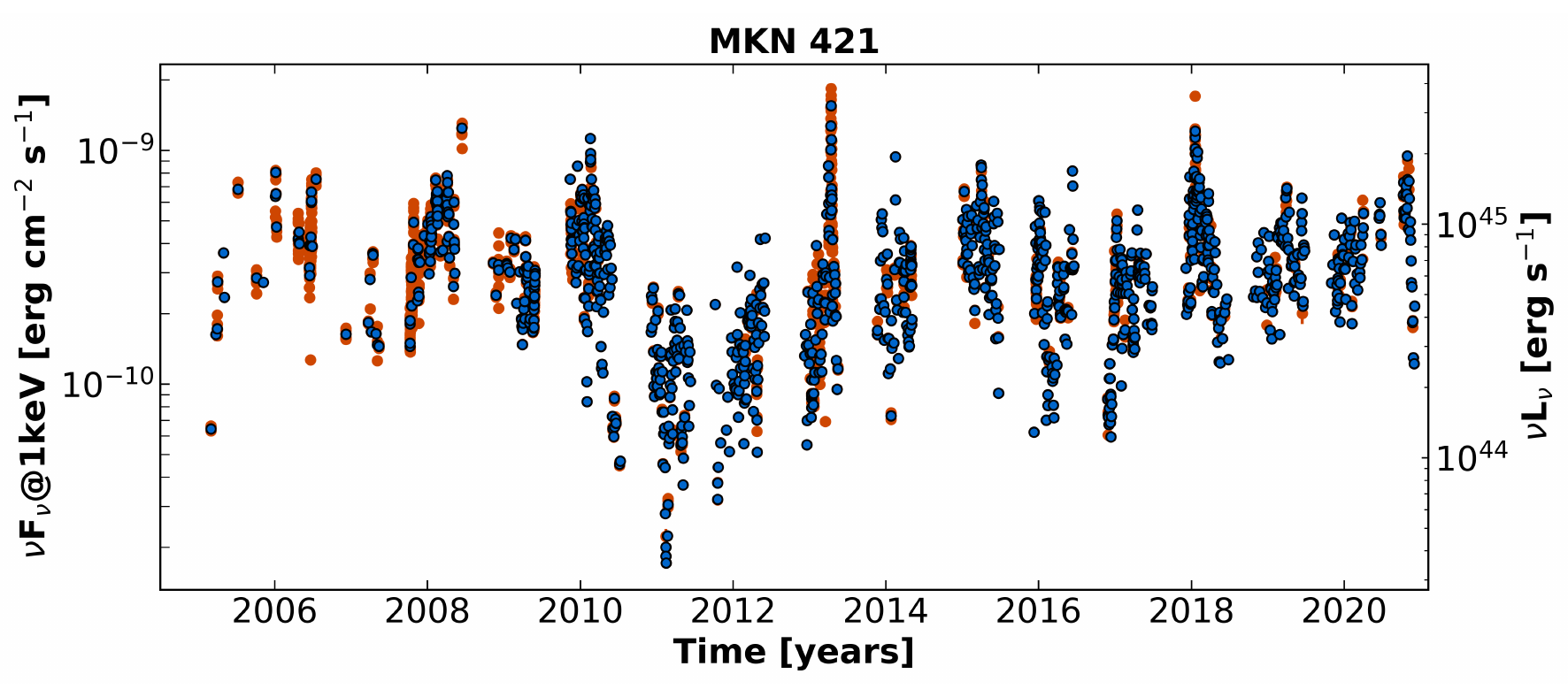}\\
\includegraphics[width=8.5cm]{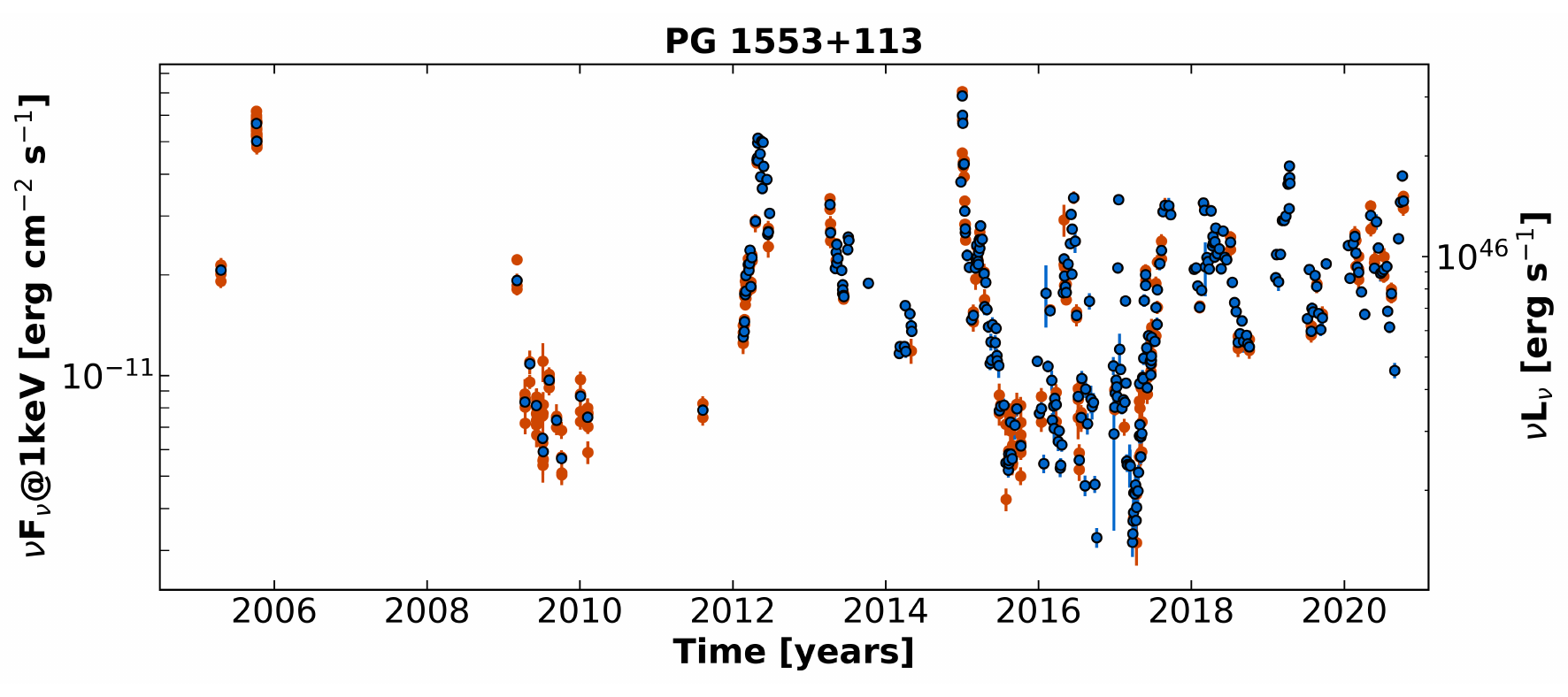}\\
\includegraphics[width=8.5cm]{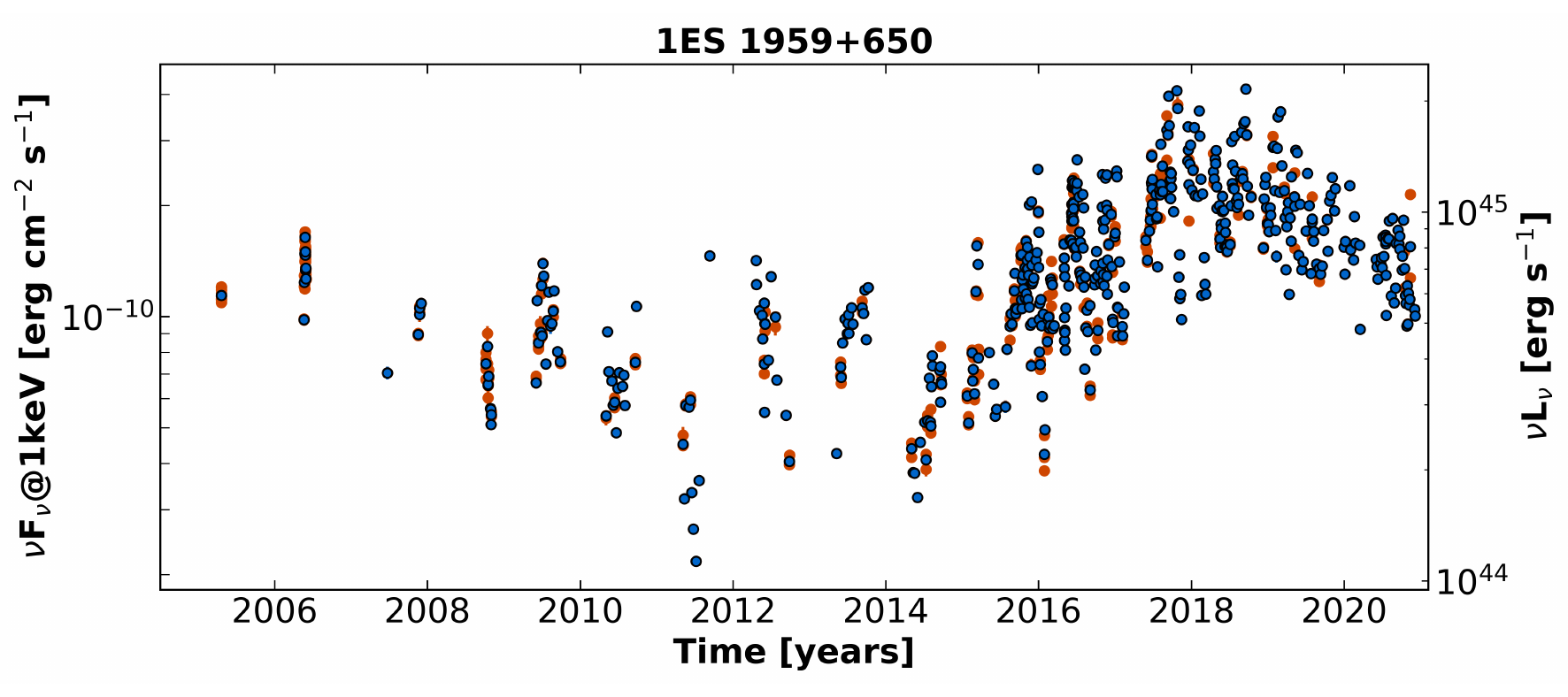} \\
\includegraphics[width=8.5cm]{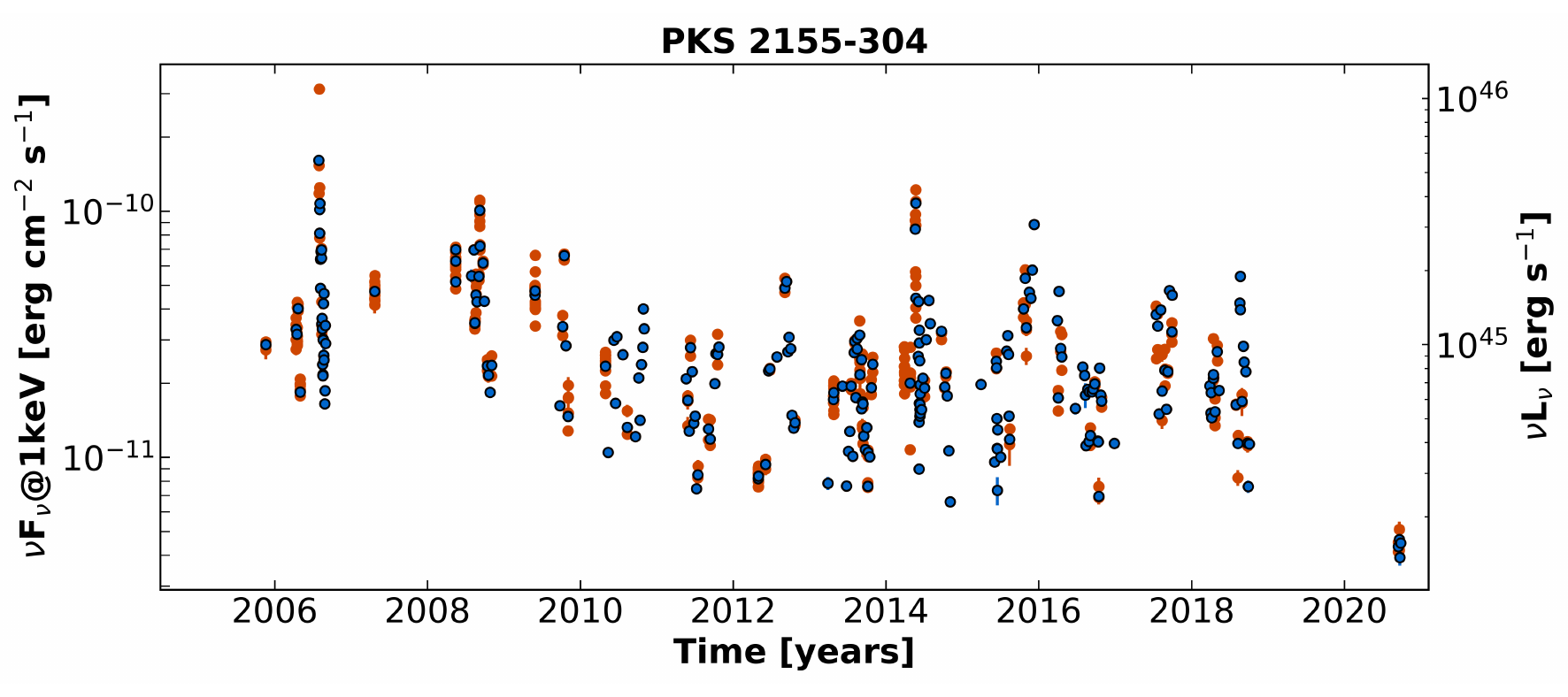}
\caption{The 1 keV light curves of HBL blazars: Mrk 501, Mrk 421, PG1553+113, 1ES1959+650, and PKS2155-304}
\label{fig:LCHBLs}       
\end{figure}

\begin{figure}
\centering
\includegraphics[width=8.5cm]{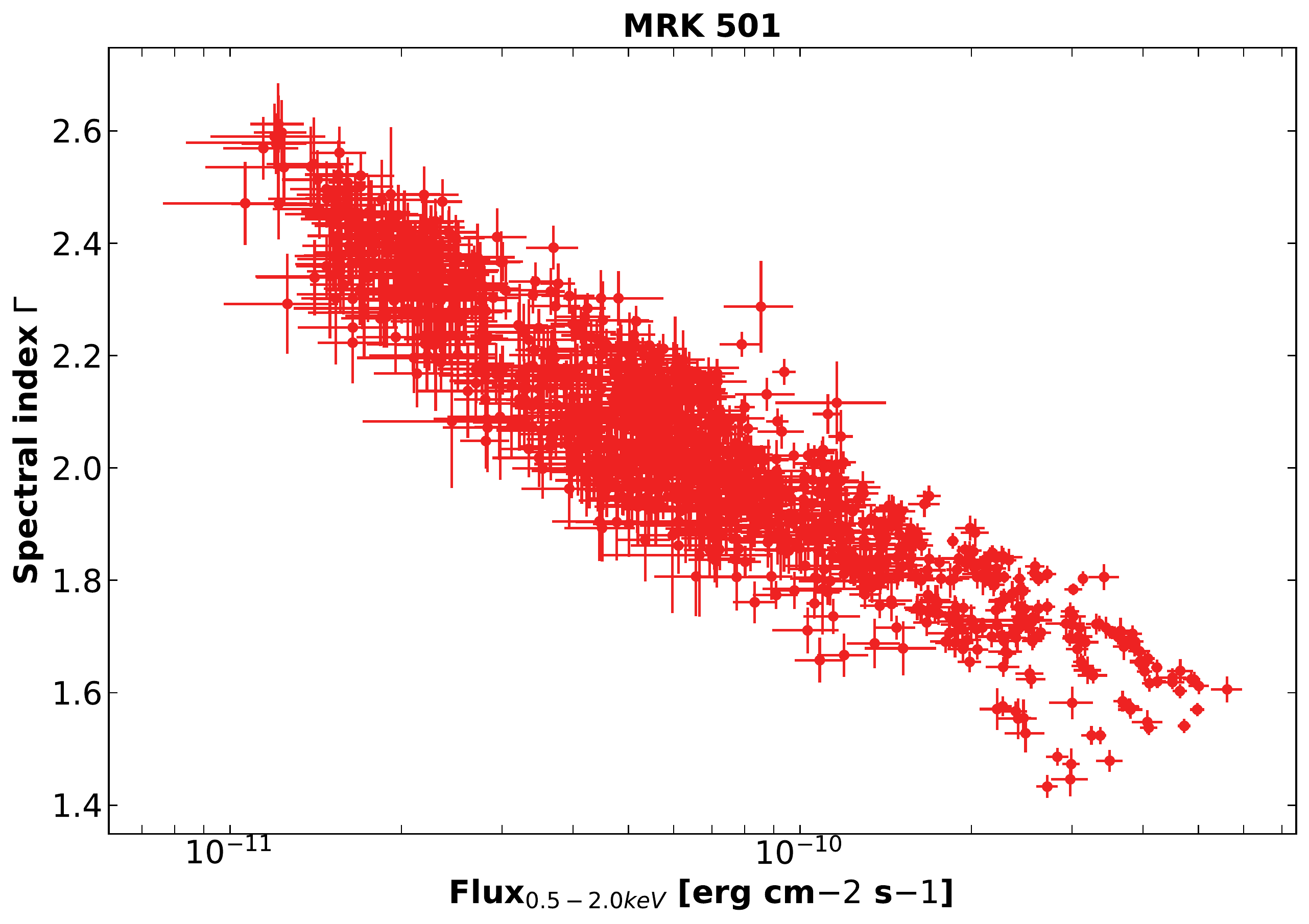}\\
\includegraphics[width=8.5cm]{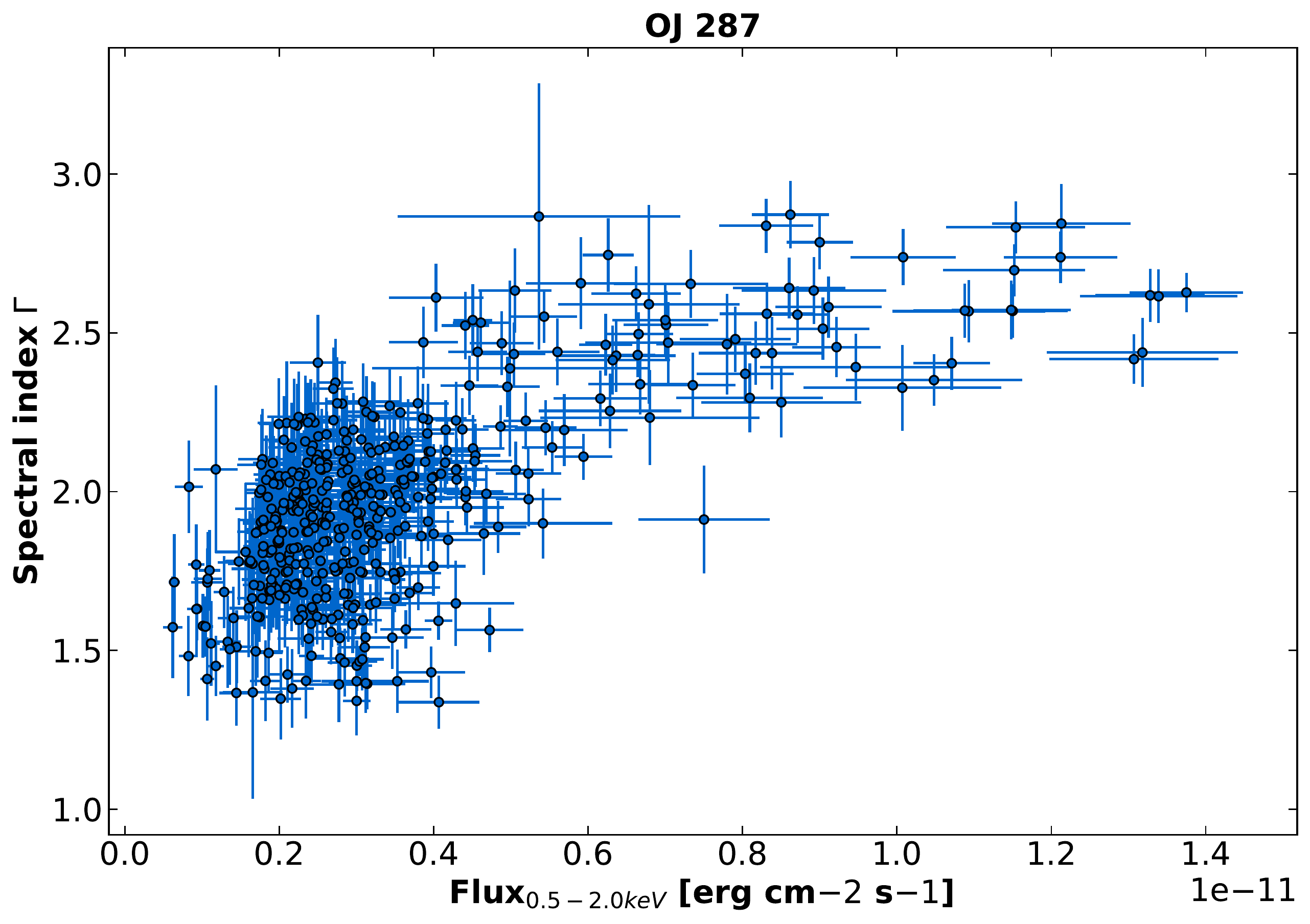} \\
\includegraphics[width=8.5cm]{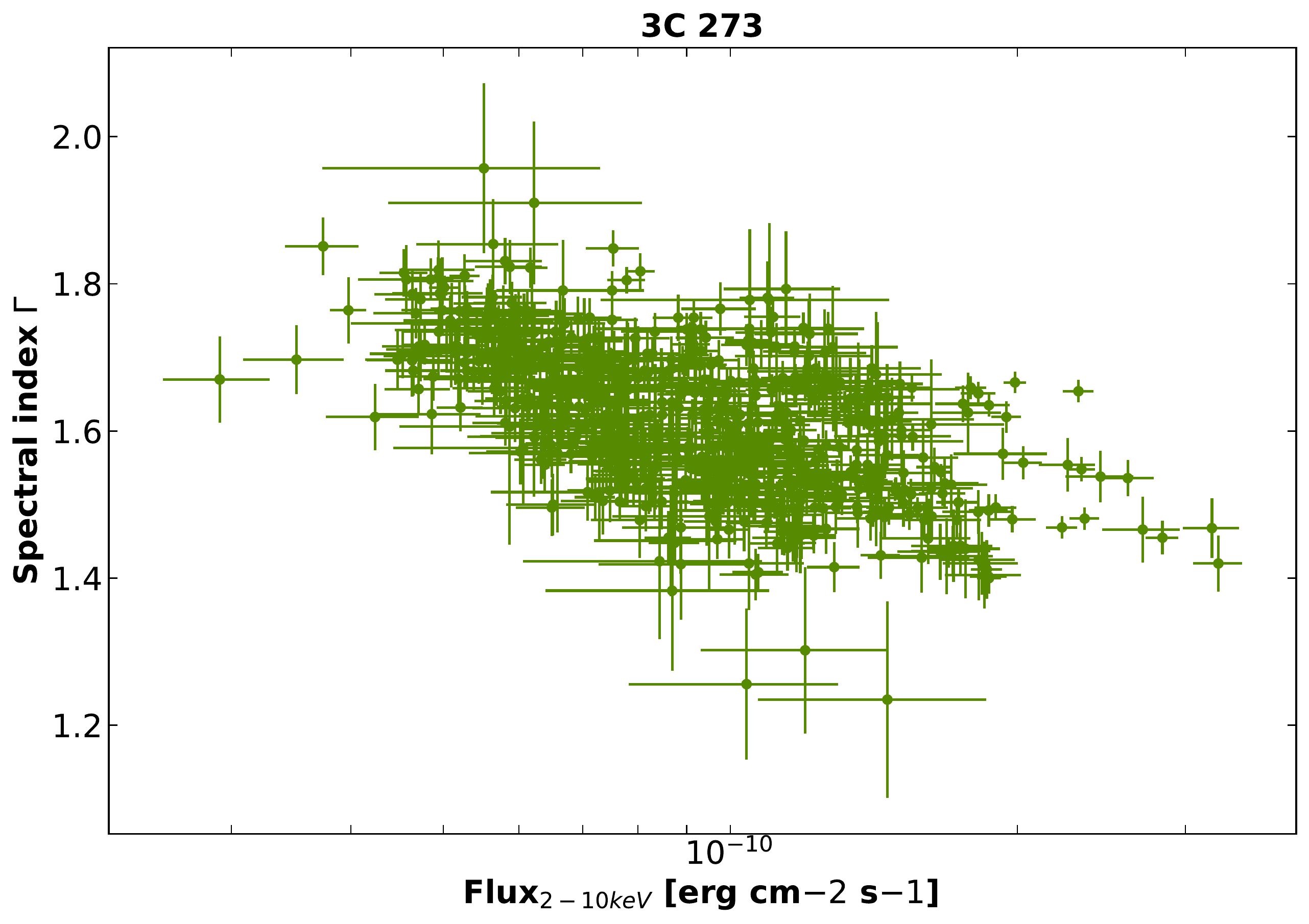}
\caption{Plot of the power-law spectral slope versus X-ray flux for the HBL blazar Mrk501 (upper panel), for the IBL object OJ287 (middle panel), and 3C273 (lower panel) illustrating the various cases of harder-when-brighter, and softer-when-brighter behaviour in blazar spectral variability.}
\label{fig:indexvsflux}       
\end{figure}

\section{Data processing}

Data reduction and scientific analysis was performed using Swift\_xrtproc, a software tool developed in cooperation between the Open Universe team and the ASI Space  Science Data Center (SSDC). This tool uses the XRT Data Analysis Software (XRTDAS\footnote{Developed under the responsibility of the ASI SSDC}), and the spectral and imaging analysis tools XSPEC and XIMAGE, included in the HEASoft package\footnote{
https://heasarc.gsfc.nasa.gov/docs/software/heasoft/}, currently released as version V6.28.

\subsection{Swift\_xrtproc}

Swift\_xrtproc executes a complete data reduction, from XRT raw data to calibrated data products. The spectral and imaging data taken in PC or WT mode are analysed following standard procedures. The main steps that are performed on each Swift-XRT observation are:

\begin{enumerate}

\item Automatic download of raw data and calibration files from one of the official Swift archives.
\item Generation of exposure maps and calibrated data products, for each snapshot
and for the entire Swift observation, using the XRTPIPELINE task and adopting standard parameters and filtering criteria.
\item Source and background spectral files generation. The source counts are estimated in a circle of 20 pixels radius when no pile-up\footnote{Pile-up occurs when more than one photon hits the same pixel in a period shorter than the XRT-CCD readout time, typically 2.5 seconds in PC mode} is present. For the case of PC mode, the background is extracted in an annular region centred around the source with radius sufficiently large to avoid contamination from source photons. For the WT mode the spectrum of the background is estimated from a deep observation taken from the XRT archive.
\item Pile-up correction. A verification of the source count-rate is carried out to determine whether the data is affected by pile-up.
In case pile-up is present the spectral data is extracted again excluding the central parts of the Point Spread Function, by taking counts in an annular region with inner radius chosen depending on the measured count-rate \citep{Vaughan06}.
\item Spectral fitting using the XSPEC package \citep{xspec} assuming a power-law and a log-parabola model.
\item Conversion of best fit spectral data to \nufnu\, units for SED plotting.
\item Photometric analysis using XIMAGE to estimate count-rates, or upper limits, in four energy bands: 0.3-1.0, 1.0-2.0, 2-10, and 0.3-10 keV, for data taken in PC readout mode.
\item Count-rate to X-ray flux conversion in the 0.5-10 keV and 2-10 keV energy bands and in \nufnu\, units at the energies of 0.5, 1.5 and 4.5 keV
\item Flux or upper limit estimation in \nufnu\, units at 1 keV either from the best fit spectrum or from the photometric data (in case the source is too weak for spectral fitting) suitable for light-curve generation and time domain analysis.
\end{enumerate}

\subsubsection{Spectral analysis}

We used the XSPEC spectral fitting package Version 12.11 to fit the X-ray data generated as described above to a power-law (eq. \ref{eq:pl}) and to a log-parabola (eq. \ref{eq:lp}) model, fixing the amount of absorbing column (NH) to the Galactic value.
\begin{equation}
N(E)=k*E^{-\Gamma}
\label{eq:pl}
\end{equation}
where $\Gamma$ is the photon index.

\begin{equation}
N(E)=k*E^{-(\alpha +\beta Log(E))}
\label{eq:lp}
\end{equation}
where $\alpha$ is the photon spectral index at 1 keV and $\beta$ is the curvature parameter.
We chose these spectral shapes because they generally provide a good description of the X-ray spectra of blazars \citep[e.g. ][]{Massaro2006} while keeping the number of free parameters to a minimum. 
Cash statistics \citep{Cash1979} was adopted for all spectral fits, grouping the data with the grppha tool to include at least one count in each energy bin \citep{Cstat2009}.

\subsubsection{Image analysis}

All the XRT observations that were carried out in PC mode were also analysed using the XIMAGE (V4.5.1) X-ray image analysis package 
\footnote{https://heasarc.gsfc.nasa.gov/xanadu/ximage/}. The procedure used is equivalent to that implemented in the Swift\_deepsky tool \citep[][]{paper1,paper2}, which performs a photometric flux 
estimation using X-ray data in four energy bands: 0.3-10 keV (full band), 0.3-1.0 keV (soft band), 1.0-2.0 keV (medium band ), and 2.0-10.0 keV (hard band). 
The image background was measured using the XIMAGE/background command, and source counts were obtained using the XIMAGE/sosta tool centring on the target and counting events in boxes whose size includes 80\% of the source flux \citep[see][for more details]{paper1}. Spectral slopes were also  estimated from the hardness ratio, defined as the ratio between the counts in the hard and soft bands. Count-rates (or upper limits) were  converted to X-ray fluxes assuming NH equal to the Galactic value, and a power-law spectrum with spectral index equal to the value estimated from the XSPEC spectral analysis when available or from the hardness ratio. 
When no spectral slope estimation was possible the spectral index was assumed to be equal to 1.8. Fluxes in \nufnu\, units, suitable for SED plotting, were calculated at the reference energies of 0.5, 1.5 and 4.5 keV. Finally, a second spectral slope estimation was obtained via a least square linear fit to the 0.5, 1.5 and 4.5 keV \nufnu\, fluxes.

\section{Results}

We processed a total of 31,068 Swift XRT observations or individual snapshots of the 65 blazars listed in Tab. \ref{TheSample}. This led to the generation of 29,050 X-ray spectra, 21,141 photometric flux estimations, and 206 upper limits.

\subsection{Best fit parameters and fluxes}

The results of the XSPEC spectral fits for every spectrum including at least 20 source counts, are reported in the on-line interactive table available at https://openuniverse.asi.it/blazars/swift, (see Fig. \ref{fig:onlinetable}). 
Column 1 gives the source name, columns 2 and 3 the J2000.0 Right ascension and Declination, columns 4 and 5 give the best fit photon spectral index with error and the reduced $\chi^2$, columns 6, 7 and 8 give the $\alpha$, $\beta$ best fit parameter and $\chi^2$ for the log-parabola model, column 9 and 10 give the observed flux in the 0.5-2.0 keV and 2-10 keV bands, column 11 gives the observation time.
The complete set of results are also available as a fits file\footnote{https://openuniverse.asi.it/OU4Blazars/oublazars\_swift\_spectra\_v1.0.fits.gz whose structure is described in Table\ref{fitstable}}.

\subsubsection{Power-law Vs Log-parabola models}

Our results confirm with large statistics that the log-parabola model generally fits best the X-ray spectral shape of HBL blazars, while LBL sources are usually best fit by a simple power-law. IBL blazars often show more complex spectra, as in these cases 
the steep end of the synchrotron component merges into the much flatter inverse Compton component in the X-ray band.
As a particular example of HBL blazar, Fig. \ref{fig:chisqdistr} plots the distribution of the best fit Cash statistics divided by the number of degrees of freedom (Cstat/d.o.f.) values for the power-law and log-parabola fits of Mrk421.
The sharper and lower value peaking red histogram for the case of log-parabola demonstrates that this spectral model is a better representation than a simple power-law. Similar results are obtained for most blazars of the HBL type. 
The X-ray spectra of IBL blazars are mostly characterised by a steep power-law spectrum  since the XRT band-pass in these objects covers the end of the synchrotron tail and occasionally show the onset of the second SED hard component. 
The X-ray spectrum of LBL blazars is well represented by a simple power-law spectrum with a flat slope, with average value of $\langle\Gamma \rangle= 1.46$. In some cases, however, (see e.g. the SED of 3C273 and CTA102 in Fig.\ref{fig:SEDsLBLs}) there is some evidence for a flattening toward the high-energy end of the XRT energy window.

The spectral shapes described above can also be noted from a simple visual inspection of the SEDs of Figs. \ref{fig:SEDsLBLs}, \ref{fig:SEDsIBLs}, and \ref{fig:SEDsHBLs} and those available in the on-line table at https://openuniverse.asi.it/blazars/swift.
\subsection{Imaging analysis}

The results of the imaging analysis, including count-rates, fluxes in the 0.5-10.0 keV and 2-10 keV bands, \nufnu\ fluxes at 0.5, 1.0, 1.5 and 4.5 keV, as well as two estimates of the spectral index, are given in the fits-formatted file mentioned above and described in  
Tab.\ref{fitstable}.

\subsection{Spectral Energy Distributions}

The best-fit spectral data and the X-ray fluxes estimated from the imaging analysis have been combined with archival multi-frequency data retrieved using the VOU-Blazars software \citep{VOU-Blazars} to build the radio to \gr\, SED of each object in the sample. The VOU-Blazars tool provides access to data from over 70 catalogues and spectral databases covering the entire electromagnetic spectrum.
Figs. \ref{fig:SEDsLBLs}, \ref{fig:SEDsIBLs}, and \ref{fig:SEDsHBLs} give examples of SEDs of representative blazars of the LBL, IBL and HBL class.  
The SED of all the blazars in the sample are available on-line at 
https://openuniverse.asi.it/blazars/swift.
Grey points represent archival data, green symbols are the XSPEC best-fit spectral data, red points are from the XIMAGE/sosta measurements, and the light yellow points are the 1 keV \nufnu\, fluxes calculated as described in section \ref{Lightcurves}. Note that the red points are not present in a high-intensity states when the WT readout mode was used, reflecting the fact that imaging analysis was only performed when the XRT was operated in PC mode.

\begin{table}
\begin{tiny}
\begin{center}
\caption{Content of the FITS file including the results of the spectral fits and of the image analysis}
\begin{tabular}{lrcl}
\hline\hline
Column & Format & Units & Description\\
\hline
Blazar name    & 15A & & Blazar name  \\
 R.A.         & D & deg & Right Ascension in degrees (J2000.0 epoch) \\ 
 Dec.                 & D & deg & Declination in degrees (J2000.0 epoch) \\
 MJD      & D & days &  Modified Julian Day of observation start time  \\
 Obs\_time & D & s & observation start time in units yyyy.ff \\
 & & & where ff is the fraction of year \\
 Sequence  & A & & Swift observation ID  \\
 Flux2\_10-PL   & D &  erg cm$^{-2}$ s$^{-1}$ & 2-10 keV flux from best fit power-law model \\
 Flux2\_10-PL\_error & D & erg cm$^{-2}$ s$^{-1}$& 1 sigma error on Flux2\_10-PL \\
 Flux05\_2-PL   & D &  erg cm$^{-2}$ s$^{-1}$ & 0.5-2.0 keV flux from best fit power-law model \\
 Flux05\_2-PL\_error & D & erg cm$^{-2}$ s$^{-1}$& 1 sigma error on Flux05\_2-PL \\
 Flux03\_3-PL   & D &  erg cm$^{-2}$ s$^{-1}$ & 0.3-3.0 keV flux from best fit power-law model \\
 Flux03\_3-PL\_error & D & erg cm$^{-2}$ s$^{-1}$& 1 sigma error on Flux03\_3-PL \\
 $\chiˆ2_{\rm r}$-PL & D & & reduced $\chiˆ2$ for power-law model \\
 d.o.f.-PL & D & & degrees of freedom for power-law model \\
 $\chiˆ2_{\rm r}$-LP & D & & reduced $\chiˆ2$ for log-parabola model \\
 d.o.f.-LP & D & & degrees of freedom for log-parabola model \\
 Photon\_index & D & & best fit photon index for power-law fit \\
 Photon\_index\_error & D & & error on best fit photon index for power-law fit \\
 Normalisation-PL & D & & best fit power-law normalisation  \\
 Normalisation-PL\_error & D & & error on best fit power-law normalisation \\
 $\alpha$ & D & & best fit  $\alpha$ parameter for log-parabola fit \\
$\alpha$\_error & D & & error on $\alpha$ parameter for log-parabola fit \\
$\beta$ & D & & best fit  $\beta$ parameter for log-parabola fit \\
$\beta$\_error & D & & error on $\beta$ parameter for log-parabola fit \\
Normalisation-LP & D & & best fit log-parabola normalisation  \\
Normalisation-LP\_error & D & & error on best fit log-parabola normalisation \\
Snapshot & 3A & & swift snapshot number \\
&&& "TOT" for full observation \\
\nufnu\,@1.0keV & D & erg cm$^{-2}$ s$^{-1}$ &  \nufnu\ flux at 1.0 keV \\
\nufnu\,@1.0keV\_error & D & erg cm$^{-2}$ s$^{-1}$ & error  \nufnu\ flux at 1.0 keV \\
Datamode  &2A & & XRT data readout mode: PC or WT \\
I-\nufnu\,@0.5keV & D & erg cm$^{-2}$ s$^{-1}$ &  \nufnu\ flux at 0.5 keV from image analysis\\
I-\nufnu\,@0.5keV\_error & D & erg cm$^{-2}$ s$^{-1}$ & error  \nufnu\ flux at 0.5 keV \\
I-\nufnu\,@1keV & D & erg cm$^{-2}$ s$^{-1}$ &  \nufnu\ flux at 1.0 keV from image analysis\\
I-\nufnu\,@1keV\_error & D & erg cm$^{-2}$ s$^{-1}$ & error  \nufnu\ flux at 1.0 keV \\
I-\nufnu\,@1.5keV & D & erg cm$^{-2}$ s$^{-1}$ &  \nufnu\ flux at 1.5 keV from image analysis\\
I-\nufnu\,@1.5keV\_error & D & erg cm$^{-2}$ s$^{-1}$ & error  \nufnu\ flux at 1.5 keV \\
I-\nufnu\,@4.5keV & D & erg cm$^{-2}$ s$^{-1}$ &  \nufnu\ flux at 4.5 keV from image analysis\\
I-\nufnu\,@4.5keV\_error & D & erg cm$^{-2}$ s$^{-1}$ & error  \nufnu\ flux at 4.5 keV \\
I-Flux0510 & D & erg cm$^{-2}$ s$^{-1}$ & 0.5-10.0 keV flux from image analysis \\
I-Flux0510\_error & D & erg cm$^{-2}$ s$^{-1}$ & error on 0.5-10.0 keV flux from image analysis \\
I-Flux210 & D & erg cm$^{-2}$ s$^{-1}$ & 2.0-10.0 keV flux from image analysis \\
I-Flux210\_error & D & erg cm$^{-2}$ s$^{-1}$ & error on 2.0-10.0 keV flux from image analysis \\
I-Photon\_index & D & &  power law photon index fit from hardness ratio\\
I-Photon\_index\_error & D & & error on photon index for power-law fit \\
SED-Photon\_index & D & &  photon index for power-law fit estimated\\
& & & from a fit to I-\nufnu\,@0.5keV, \\
& & & I-\nufnu\,@1.5keV and I-\nufnu\,@4.5keV\\
SED-Photon\_index\_error & D & &  error on photon index \\
SED-Photon\_index\_error & D & & error on photon index for power-law fit \\
CR(0.3-10keV)  & D &  cts s$^{-1}$ & count-rate in 0.3-10.0 keV energy band \\
CR(0.3-10keV)\_error  & D &  cts s$^{-1}$ & error on count-rate in 0.3-10.0 keV energy band \\
O-\nufnu\,@1keV & D & erg cm$^{-2}$ s$^{-1}$ &overall \nufnu\ flux at 1.0 keV  for lightcurves\\
O-\nufnu\,@1keV\_error & D & erg cm$^{-2}$ s$^{-1}$ &error on overall \nufnu\ flux at 1.0 keV \\
Upper limit flag & 2A &  & flag set to 'UL' if the count rate is a 3$\sigma$ limit, \\
& & & blank otherwise\\
\hline\hline
\end{tabular}
\label{fitstable}
\end{center}
\end{tiny}
\end{table}

\subsubsection{Synchrotron peak energy (\nupeak\,)}

Reliable values of \nupeak\, can be estimated directly from the X-ray spectrum in all sources where the log-parabola model is a good fit to the data and \nupeak\, is located inside or close to the XRT band-pass. From \cite{Massaro2006} we have 
\begin{equation}
\nu_{\rm peak} = 10^{(2-\alpha)/2\beta}
\label{eq:ep}
\end{equation}
where $\alpha$ and $\beta$ are the parameters of eq. \ref{eq:lp}.
We have calculated \nupeak\, in all cases where well estimated values of $\alpha$ and $\beta$ could be obtained and eq. \ref{eq:ep} gives values that are close or within the Swift XRT X-ray band. We found that large variations of this parameter are detected in many objects. As an example of this behaviour Fig.\ref{fig:mkn421-epeak} plots \nupeak\, as a function of time for the case of Mrk421, the object most frequently observed in the sample, showing frequent large changes spanning over two orders of magnitude, from below 0.1 to over 20 keV.
Large \nupeak\, changes are common in most HBL blazars, as illustrated in Fig.\ref{fig:nupeak_distr} which plots the \nupeak\, distribution of Mrk421 together with three other representative sources. 
Values close to the low and high energy ends of Fig.\ref{fig:nupeak_distr} should be taken with caution \citep[and perhaps treated as limits, see e.g.][]{2016MNRAS.461L..26K,2020ApJS..247...27K} since they are at the edge of the XRT energy range. However, the
frequency of occurrence is low and the impact on the overall shape of the distribution is limited.

\subsection{Time domain data}\label{Lightcurves}

Detailed temporal studies of the X-ray emission in blazars is beyond the scope of this paper. In the following we limit ourselves to present the 1 keV light-curves of all the blazars in the sample and briefly comment on the long-term variability and flux variations between neighbouring snapshot, based on visual inspection of the data. 
A comprehensive study of blazars temporal behaviour based on the data presented here will be the subject of a future publication.

The lightcurves presented in this paper are based on 1 keV (\nufnu\,) fluxes, calculated as the best fit value (and statistical error) of the power-law normalisation, multiplied by 1.6$\times 10^{-9}$, to convert from XSPEC units to erg cm$^{-2}$ s$^{-1}$ , or converting to 1 keV the XIMAGE count-rate estimation in the soft band, when less than 20 source counts were detected, or 3$\sigma$ upper limits estimated with the XIMAGE/sosta command in case of non-detection.

\subsubsection{Long-term lightcurves}

1 KeV \nufnu\, lightcurves covering the period November 2004 to the end of 2020 of all the sources in the sample can be accessed 
on-line at https://openuniverse.asi.it/blazars/swift/. Figs. \ref{fig:LCLBLs}, \ref{fig:LCIBLs}, and \ref{fig:LCHBLs} show examples of lightcurves of 5 representative LBL, IBL, and HBL blazars, respectively.  
Blue points represent fluxes averaged over the entire observation, red points correspond to single snapshots.
Large luminosity variability on different amplitudes and timescales, ranging from  factor of a few to over a factor 200, is present in all sources. The ratio between the maximum and minimum observed flux in each source is listed in column (5) of table \ref{TheSample}.

Lightcurves based on integrated XRT fluxes in the 0.5-2.0 keV, 0.3-3.0 keV, 0.5-10 keV, and 2-10 keV bands, or \nufnu\, fluxes at the energy of 0.5 keV, 1.5 keV, 4.5 keV, can be produced using the data provided in the fits file described in Tab. \ref{fitstable}.

\subsubsection{Harder-when-brighter and softer-when-brighter behaviour}

Spectral changes correlated with intensity variations in blazars were first noticed in early EXOSAT X-ray observations of PKS2155-304 \citep{Morini1986} and of Mrk421 \citep{George1988}. A study conducted on a sample of 36 BL Lacs showed that this behaviour is a common feature of BL Lacs objects \citep{Giommi1990}.
In the following we investigate these correlations in our  sample, which includes all types of blazars and a much larger number of observations.

The upper panel of Fig.\ref{fig:indexvsflux} plots the best fit power-law spectral index as a function of X-ray flux for the source Mrk501. A clear trend is present, with steep spectral slopes ($\Gamma \gsim$ 2.5) being observed during low flux periods and much harder ($\Gamma \sim$ 1.5) values during high-intensity states. This harder-when-brighter behaviour is  common to the HBL sources in our sample as can also be noticed in the SEDs of Fig. \ref{fig:SEDsHBLs}.

A case of the opposite (softer-when-brighter) trend is shown in the middle panel of Fig. \ref{fig:indexvsflux} for the case of the IBL blazar OJ287. In this intermediate \nupeak\, situation the X-ray flux strongly increases when \nupeak\, reaches values larger than 
$\approx 10^{14.5}$ Hz and the steep tail of the synchrotron SED component enters the X-ray band. When \nupeak\, is instead located at lower frequencies, the synchrotron component does not reach X-ray energies and the flux in the X-ray band is only due to the hard inverse Compton component.

The lower part of Fig. \ref{fig:indexvsflux} illustrates the case of the LBL object 3C273, which shows an apparent  harder-when-brighter behaviour. This is likely not due to an intrinsic hardening of the inverse Compton emission from the jet but rather to a changing mix of different components. This is because during high intensity states the flat radiation from the jet dominates the steeper X-ray emission from the accretion onto the supermassive black hole \citep[e.g.][]{grandipalumbo}. 

\subsubsection{Rapid variability}

Large X-ray luminosity variability on a variety of timescales in blazars is common and amply documented in the literature \citep[e.g.][and references therein]{Giommi1990,2016ApJ...831..102K,2018ApJ...867...68W,2020ApJS..247...27K,2018ApJ...858...68K, 2018ApJ...854...66K, 2017ApJ...848..103K, 2018ApJS..238...13K, 2018MNRAS.473.2542K}.
We searched for large amplitude ($\gsim$ 50\%) variability on timescales of the order of one or a few hours by comparing the flux observed during contiguous or neighbouring snapshots, which are typically separated by one Swift orbit, or $\sim $ 1.5 hours.
Large luminosity variations on these timescales are rarely observed in our sample. This may reflect the fact that this type of variations are intrinsically rare in blazars, but it could also be due to limited sampling or observing strategy, as in many cases the observations consisted of single snapshot exposures.
The list of the thirteen objects where rapid variability has been noticed is given in Table \ref{Fastvariableblazars} where column 1 gives the blazar name, column 2 gives the Swift Observation ID, column 3 the time in MJD units, column 4 the time difference between the snapshots where the variability was observed, and column 5 gives the ratio between the flux measured in the last snapshot and that of the first snapshot.

\section{Discussion}

The analysis presented in the previous sections shows that long-term large amplitude X-ray variability 
is present in the data of all blazars of the sample on a variety of timescales (see table \ref{TheSample} and the lightcurves of Figs. \ref{fig:LCHBLs}, \ref{fig:LCIBLs} and \ref{fig:LCLBLs}). 

No obvious periodicities or regularities can be seen from a visual inspection of the lightcurve of any object. However, these results cannot be considered an unbiased view of the long-term behaviour of blazars since X-ray sources that are  frequently observed by Swift are often pointed as Target of Opportunity triggered by flaring activity discovered in other parts of the electromagnetic spectrum.
Nevertheless, the results presented here provide a new rich dataset that is very useful to study the observational properties of blazars and constrain physical models in the energy and time domain. In the following we make some initial considerations on blazars properties that are not significantly affected by the selection biases of the sample.

\subsection{X-ray emission in blazars of different types : HBLs vs IBLs vs LBLs} 
Our sample includes 24 HBL, 12 IBL, and 29 LBL blazars. 
The X-ray band is particularly important in the energy spectrum blazars of all types, as the X-ray flux of HBL sources is entirely due to synchrotron emission, while in IBL objects the end of the synchrotron tail often merges with the second (inverse Compton or other) high-energy component. In LBL sources the X-ray flux instead maps the beginning of the second SED bump. Different spectral shapes, variability properties, and probably also time lags between the soft and hard X-ray band, are to be expected in blazars of different types. 

The location and variability properties of \nupeak\, provide crucial information about the physical conditions in blazars jets. Among the important physical parameter that can be constrained are the maximum particles energy, the acceleration and cooling times, the size of the emitting region, and the distribution of the particles responsible for the non-thermal radiation that we observe.
The results obtained with our large data-set on this important parameter can be summarised as follows:
\begin{itemize}
    \item The SEDs of the 65 objects in our sample,
    some of which are shown in Figs. \ref{fig:SEDsLBLs}, \ref{fig:SEDsIBLs}, \ref{fig:SEDsHBLs}, confirm with large statistics that \nupeak\, in blazars ranges from $\sim 10^{12}$ Hz, to well over $10^{18}$ Hz.
    \item  The position of \nupeak\, is not constant in time, as noticed over twenty years ago in the BeppoSAX data  of Mrk501 and 1ES2344+514 \citep{pian1998,giommi2000}, and more recently studied in detail in some objects \citep[e.g.][]{2018MNRAS.473.2542K,2020ApJS..247...27K}.
    Our results show that frequent and large variations of \nupeak , spanning a range of well over a factor 100 in some objects, is a very common, likely ubiquitous, feature of HBLs.
    As an example, Fig. \ref{fig:mkn421-epeak} illustrates how frequently \nupeak\, changes in time and amplitude in Mrk421, the blazar most frequently observed by Swift.
    The SEDs plotted in Fig \ref{fig:SEDsHBLs}, and those available on-line show that this phenomenon is present in most HBLs of the sample. 
    Fig. \ref{fig:nupeak_distr} shows the distribution of
    \nupeak\, in four representative HBL blazars, where we can see that blazars of this type spend a significant fraction of the time with \nupeak\, values that are more than a factor 5 away from the average value. 
    Note that these histograms may be a partially distorted representation of the intrinsic probability distributions because of the selection biases of the sample. This is because ToO observations tend to follow large \nupeak\, values  during flares, rather than quiet periods. In addition, the relatively narrow band-pass of XRT limits the range over which \nupeak\, can be safely estimated to $\sim 0.5$ keV and 8-10 keV.
    
    \item Variations of  \nupeak\, in IBL sources are difficult to detect with XRT data alone since \nupeak\, in these objects is located outside the X-ray band. Despite that, \cite{Giommi2008} combining Swift UVOT and XRT data, found evidence for \nupeak\ variations in the IBL blazar S50716+714. Harder-when-brighter trends has been found in the infrared and optical part of the spectrum of the same object by \cite{Xiong} and \cite{Morokuma2020}, suggesting that large \nupeak\, variations should be present in IBL blazars as well.
        
    \item Variations of the SED peak frequency in LBLs are also hard to observe because of the lack of multi-epoch observations in the far infrared, where \nupeak\ is located in these sources. However, at least in the cases of CTA 102, 3C279 and 3C454.3 (see fig. \ref{fig:SEDsLBLs}), where large intensity variations in the IR/optical/UV and X-ray band have been detected, no significant spectral changes have been detected, consistently with no evidence for a large increase of \nupeak\, in LBLs. 
    
    An optical-infrared monitoring of $\gamma$-ray emitting blazars of different types \citep{Bonning} detected a tendency for FSRQs (all of which are of the LBL type) to become bluer when fainter. This apparent hardening was interpreted as due to the blue accretion disk emission becoming increasing dominant during faint states of the jet component. When the sources were found in a bright state the spectrum tended to become redder (steeper), implying that  \nupeak\, in these sources was not moving from the far IR to energies close to the optical band. 
    
    \item Most of the X-ray flux enhancements in HBL sources is due to the effects of spectral hardening, rather than to an overall flux increase at all energies. This effect is larger when \nupeak\, is located in the very soft X-ray band. In this situation a small increase of this parameter flattens the X-ray spectrum inducing large flux increases.
    On the other hand, when \nupeak\, reaches values that are larger than the energy band where the flux is measured, no significant flux increase is often observed. This behaviour may be due to a \nupeak\, increase without an associated increase in the normalisation of the spectrum, or to the onset of a second hard component related to a different emitting region. This second hypothesis was first considered by \cite{giommi2000}, who interpreted the \nupeak\, shift in 1ES2344+514 as the result of a different component emerging at high energies.
\end{itemize}

Most values of \nupeak\, reported in the literature are estimated 
under one of these circumstances: a) on the basis of a single X-ray observation, b) based on the average value of several X-ray measurements, or c) on sparse archival X-ray data taken at different epochs. In all cases the value of \nupeak\, may be biased for the following reasons:
 when a blazar has been observed many times, the observations are often the result of monitoring campaigns carried out in response to the announcement of a large flare or of a high state of the object. Since there is a strong correlation between source intensity and \nupeak\,\citep[in most cases higher \nupeak\ when the source brightens, e.g.][]{2018MNRAS.473.2542K,2020ApJS..247...27K} the average flux and the corresponding \nupeak\, will be biased towards high values.
 When instead a blazar has only one X-ray observation (e.g. objects discovered in the RASS survey, \cite{Voges1999}, and never observed afterwards), the value of \nupeak\, is a random draw from the (intrinsic, perhaps time dependent) \nupeak\, distribution similar to those shown in Fig. \ref{fig:nupeak_distr}. If this single draw is a low value that corresponds to a flux below the X-ray sensitivity limit, the blazar is not even discovered. Those above the X-ray limit but close to it, as in many real cases, will have a value of \nupeak\ that is biased towards high values.
 
 Some of the so-called extreme blazars \citep{Biteau2000} might therefore not be really extreme sources, but objects that spend most of their time with moderately high \nupeak\, values and happened to have been discovered in a high state.

 \subsection{Unexpected findings and future samples}
 
 The existence of transient blazars was not known until the detection of 4FGLJ1544.3-0649 \citep{TransientBlazar}, one of the objects in our sample.
 This object remained hidden in the archives as one of the many anonymous radio sources with no associated high-energy emission, until May 2017 when \cite{ciprini2017} reported the emergence of a new Fermi-LAT transient source, triggering observations at other frequencies. These new data led to the identification of a new blazar, which for a few months became one of the brightest objects of this type in the X-ray and \gr\, bands. 
 
 The unexpected existence of
 transient blazars could complicate or even modify our perception of this type of sources if 4FGLJ1544.3-0649 does not represent an isolated case but rather the tip of the iceberg of a previously unnoticed phenomenon.
 It is likely that similar sources are present in the existing archival data or will be found in future sensitive observations. 
 
 The SRG/eROSITA \citep{e-rositabook,e-rosita2020} X-ray telescope is conducting an all-sky X-ray survey with good sensitivity, repeated at different epochs. This survey will have a deep impact on blazar science with the discovery of a few thousand new objects with multiple detections, especially of those located near the ecliptic poles, thus mitigating or largely removing the biases described above. This extremely valuable data-set, alongside the science archives of the forthcoming and previous space missions and ground based observatories, constitute a colossal, ever growing, reservoir of information. It is clear that it is not possible to efficiently convert this enormous potential into knowledge following a traditional approach based on the analysis of low or intermediate level data from every instrument, each requiring specific expertise.
 Only the easily-accessible/transparent availability of science-ready data products will enable us to fully exploit the great potential of open archives, giving us, in this particular case, a much more complete view of blazars. The work presented in this paper aims at being a step in this direction.

\section{Conclusion}\label{conclusion}

We have analysed the Swift XRT data taken in PC and WT modes of 65 blazars observed at least 50 times from launch in late 2004 to the end of 2020. We have processed a total of 31,068 Swift XRT observations and individual snapshots using a software pipeline called Swift\_xrtproc that automatically retrieves the data, performs a spectral and photometric analysis, and writes the results to an archive database. This work complements and extends the results presented in \cite{paper1} based on a X-ray imaging analysis of the blazars observed by Swift in PC mode. It represents another step towards the development of a multi-messenger multi-temporal high-transparency archive of blazars data products and results.
We have used this unprecedented dataset to describe the general temporal and spectral behaviour of blazars of different types. Specific papers on timing analysis, spectral modeling, and neutrino emission prediction will be presented in the near future. 

\section*{Data Availability}
All results generated as part of this work are available through the tools and web pages of the Open Universe platform, e.g. the VOU-BLazars application \citep{VOU-Blazars} and the interactive table at https://openuniverse.asi.it/blazars/swift/, the ASI SSDC, the Virtual Observatory, and as a fits formatted file that can be downloaded at the following URL: \\
{\footnotesize https://openuniverse.asi.it/OU4Blazars/oublazars\_swift\_spectra\_v1.0.fits.gz}

\begin{table}[ht]
\begin{footnotesize}
\begin{center}
\caption{Blazars where flux variability was observed between consecutive or neighbouring snapshots }
\begin{tabular}{lcccc}
\hline\hline
Name  & Obsid & Date & $\Delta$ T & f$_{t2}$/f$_{t1}$ \\
      &       & (MJD)  & (Ks) & \\
(1) & (2) & (3) & (4) & (5) \\
\hline
1H0323+342  & 00035372001 & 53922.8 & 5 & 2.4 \\
1H0323+342 & 00035630002 & 53926.6 & 6 & 1.7 \\
1ES0414+009 & 00081691001 & 57353.9 & 6 & 0.5\\
TXS0506+056 & 00083368006 & 58031.7 & 5 & 0.3  \\
3C120  & 00037594042 & 56934.2 & 5 &  1.5\\
OJ287 & 00033756119 & 57792.3 & 11 & 0.3 \\
OJ231 & 00031219001 & 54624.9 & 15 & 4.1 \\
PKS0548-322 & 00044002081 & 58370.9 & 5 & 0.3 \\
Mrk421 &00030352011 & 53905.4 & 5 & 2.9 \\
PG1553+113 & 00031368076 & 57021.8 & 5 & 0.7\\
EXO1811.7+3143  & 00013748001 & 59125.5 & 4 & 1.5 \\
1ES1959+650  & 00034588249 & 59164.9 & 1 & 1.7 \\
PKS2155-304 &00030795001 & 53946.1 & 75 & 2.7 \\
BLLac    &	00034748002 & 57668.3 & 5& 0.2 \\
\hline\hline
\end{tabular}
\label{Fastvariableblazars}
\end{center}
\end{footnotesize}
\end{table}

\section*{Acknowledgements}

We acknowledge the use of data, analysis tools and services from the Open Universe platform, the ASI Space Science Data Center (SSDC), the Astrophysics Science Archive Research Center (HEASARC), the Astrophysics Data System (ADS), and the National Extra-galactic Database (NED).\\
\textbf{PG} acknowledges the support of the Technische Universit\"at M\"unchen - Institute for Advanced Studies, funded by the German Excellence Initiative (and the European Union Seventh Framework Programme under grant agreement no. 291763) and the support by the Excellence Cluster ORIGINS, which is funded by the Deutsche Forschungsgemeinschaft (DFG, German Research Foundation) under Germany's Excellence Strategy -EXEC-2094 - 390783311.\\
\textbf{CHB} acknowledges the support of ICRANet and the Brazilian government, funded by the CAPES Foundation, Ministry of Education of Brazil under the project BEX 15113-13-2.\\
\textbf{UBdA} acknowledges the support of a CNPq Productivity Research Grant no. 311997/2019-8 and a Serrapilheira Institute Grant number Serra - 1812-26906. He also acknowledges the receipt of a FAPERJ Young Scientist Fellowship no. E-26/202.818/2019.\\
\textbf{RM} acknowledges the financial support of INAF (Istituto Nazionale di Astrofisica), Osservatorio Astronomico di Roma, ASI (Agenzia Spaziale Italiana) under contract to INAF: ASI 2014-049-R.0 dedicated to SSDC. \\
\textbf{NS} acknowledges the support by the Science Committee of RA, in the frames of the research project No 20TTCG-1C015.\\
We thank the referee for useful comments.



\bibliographystyle{mnras}
\bibliography{SwiftSpectraMNRAS4arXiv} 



\bsp	
\label{lastpage}
\end{document}